%% file: math7.tex
\documentclass[12pt]{amsart}
\usepackage[T1]{fontenc}
\usepackage{amssymb}
\usepackage{a4wide}
\usepackage{graphicx}
\usepackage{verbatim}
\usepackage{amsmath}
\usepackage{version}
\usepackage{yhmath}

 \theoremstyle{plain}    
 \newtheorem{thm}{Theorem}%%[section]
 \numberwithin{equation}{section} %% Comment out for sequentially-numbered
 \theoremstyle{plain}
 \newtheorem{prop}{Proposition} %%Delete [thm] to re-start numbering
 \theoremstyle{plain}    
 \theoremstyle{plain}    
 \newtheorem{cor}{Corollary} %%Delete [thm] to re-start numbering
 \theoremstyle{plain}    
 \newtheorem{conj}{Conjecture} %%Delete [thm] to re-start numbering
 \theoremstyle{plain}    
 \newtheorem{question}{Question} %%Delete [thm] to re-start numbering
\theoremstyle{definition}
\newtheorem{rem}{Remark} %%Delete [thm] to re-start numbering
\theoremstyle{definition}
\newtheorem{Def}{Axioms} %%Delete [thm] to re-start numbering

\begin{document}
\input{def.tex}

\title%[Resonant eigenstates in quantum chaotic scattering ]
{Resonant eigenstates for a quantized chaotic system}

\author[S. Nonnenmacher]{St\'ephane Nonnenmacher}
\author[Mathieu Rubin]{Mathieu Rubin}
\address{Service de Physique Th\'eorique, 
CEA/DSM/PhT, Unit\'e de recherche associ\'e CNRS,
CEA/Saclay,
91191 Gif-sur-Yvette, France}
\email{snonnenmacher@cea.fr}

\begin{abstract}
We study the spectrum of quantized open maps, 
as a model for the resonance spectrum of quantum
scattering systems. We are particularly interested in open
maps admitting a fractal repeller. Using the ``open baker's map'' 
as an example, we numerically
investigate the exponent appearing in the Fractal Weyl law
for the density of resonances; we show that this exponent is not 
related with the ``information dimension'', but rather the Hausdorff
dimension of the repeller. We then 
consider the {\em semiclassical measures} associated with 
the eigenstates: we prove that these measures are
conditionally invariant with respect to the classical dynamics.
We then address the problem
of classifying semiclassical measures among conditionally invariant ones. 
For a solvable model, the ``Walsh-quantized'' open baker's map, we manage to
exhibit a family of semiclassical measures with simple self-similar properties.
\end{abstract}

\maketitle

%%%%%%%%%%%%%%%%%%%%%%
%%%%%%%%%%%%%%%%%%%%%%
\section{Introduction}
%%%%%%%%%%%%%%%%%%%%%%
%%%%%%%%%%%%%%%%%%%%%%

%%%%%%%%%%%%%%%%%%%%%%
\subsection{Quantum scattering on $\IR^D$ and resonances} 
%%%%%%%%%%%%%%%%%%%%%%

In a typical scattering system, particles of positive energy
come from infinity, interact with a localized potential $V(q)$, and then leave
to infinity. The corresponding  quantum Hamiltonian 
$H_\hbar=-\hbar^2\Delta +V(q)$ has an
absolutely continuous spectrum on the positive axis.
Yet, the Green's function $G(z;q',q)=\la q'|(H_\hbar-z)^{-1}|q\ra$ admits
a meromorphic continuation from the upper half plane $\set{\Im z>0}$ to (some
part of) the lower
half-plane $\set{\Im z<0}$. This continuation generally has 
{\it poles} $z_j=E_j-\i\Gamma_j/2$, $\Gamma_j>0$,
which are called {\it resonances} of the scattering system.

The probability density of the corresponding ``eigenfunction'' $\varphi_j(q)$
decays in time like $\e^{-t \Gamma_j/\hbar}$, so physically $\varphi_j$ represents
a metastable state with decay rate $\Gamma_j/\hbar$, or lifetime $\tau_j=\hbar/\Gamma_j$.
In the semiclassical limit $\hbar\to 0$, we will call ``long-living''
the resonances $z_j$ such that $\Gamma_j=\cO(\hbar)$, equivalently with lifetimes 
bounded away from zero.

The eigenfunction $\varphi_j(q)$ is meaningful only
near the interaction region, while its behaviour outside that region
(exponentially increasing outgoing waves) is clearly unphysical.
As a result, one practical method to compute resonances (at least approximately)
consists in adding a {\it smooth absorbing potential} $-\i W(q)$ to the Hamiltonian
$\hat H$,
thereby obtaining a nonselfadjoint operator $H_{W,\hbar}=H_\hbar-\i W(q)$.
The potential $W(q)$ is supposed to vanish
in the interaction region, but is positive outside: its effect is to {\em absorb}
outgoing waves, as opposed to a real positive potential which would reflect
the waves back into the interaction region. Equivalently, the (nonunitary)
propagator $\e^{-\i H_{W,\hbar}/\hbar}$ kills wavepackets localized oustide the
interaction region. 

The spectrum of $H_{W,\hbar}$ in some neighbourhood of the
positive axis is then made of discrete eigenvalues $\tilde{z}_j$ associated with 
square-integrable eigenfunctions $\tilde\varphi_j$. 
Absorbing Hamiltonians of the type of $H_{W,\hbar}$ have been widely used in
quantum chemistry to study reaction or dissociation dynamics
\cite{LeforWyatt83,SeidMill92}; in those works it is implicitly assumed that eigenvalues
$\tilde{z}_j$ close to the real axis are small perturbations of the resonances
$z_j$, and that the corresponding eigenfunctions $\varphi_j(q)$, $\tilde\varphi_j(q)$ 
are close to one another in the interaction region.
Very close to the real axis (namely, for $|\Im\tilde{z}_j|=\cO(\hbar^n)$
with $n$ sufficiently large), one can prove that this is indeed the case \cite{Stef05}.
Such very long-living resonances are possible when the classical dynamics admits a
trapped region of positive Liouville volume.
In that case, resonances and the associated eigenfunctions can be approximated
by quasimodes of an associated closed system \cite{TangZw98}. 

%%%%%%%%%%%%%%%%%%%%%%%%%%%%%%%%%%%%%%%%%%%
\subsection{Resonances in chaotic scattering} 
%%%%%%%%%%%%%%%%%%%%%%%%%%%%%%%%%%%%%%%%%%%

We will be interested in a different situation, where the set of 
trapped trajectories has volume zero, and is a fractal hyperbolic repeller. 
This case encompasses the
famous 3-disk scatterer in 2 dimensions \cite{GaspRice89}, or its smoothing,
namely the 3-bump
potential introduced in \cite{Sjo90} and numerically studied in \cite{Lin02}.
Resonances
then lie deeper below the real line (typically, $\Gamma_j\gtrsim \hbar$),
and are not perturbations of an associated real spectrum. 
Previous studies have focussed in counting the number of resonances in small
disks around the energy $E$, in the semiclassical r\'egime. Based on the
seminal work of Sj\"ostrand \cite{Sjo90}, several authors have conjectured
the following Weyl-type law:
\bequ\label{e:Weyl-law}
\#\set{z_j\in \Res(H_\hbar)\,:\,|E-z_j|\leq \gamma\hbar}\sim C(\gamma)\hbar^{-d}\,.
\end{equation}
Here the exponent $d$ is related to the trapped set at energy $E$: the latter has
(Minkowski) dimension $2d+1$. 
This asymptotics was numerically checked by Lin and Zworski for
the 3-bump potential
\cite{Lin02,LSZ03}, and by 
Guillop\'e, Lin and Zworski for scattering on a hyperbolic surface \cite{GLZ04}. 
However, only the upper bounds for the number
of resonances could be rigorously proven \cite{Sjo90,Zw99,GLZ04,SjoZw05}. 

To avoid the
complexity of ``realistic'' scattering systems,
one can study simpler models, namely quantized open maps on a compact phase space, for instance
the quantized open baker's map studied in \cite{NonZw05,NonZw06,KNPS06} (see \S\ref{e:open-map}). 
Such a model is meant to mimick
the propagator of the nonselfadjoint Hamiltonian $H_{W,\hbar}$, in the case where the
classical flow at energy $E$ is chaotic in the interacting region. The above
fractal Weyl law has a direct counterpart in this setting; such a fractal scaling
was checked for the open kicked rotator in \cite{SchoTwo04}, 
and for the ``symmetric'' baker's map in \cite{NonZw05}.
In \S\ref{s:infodim?} we numerically check this fractal law for an 
asymmetric version of the open baker's map; apart from
extending the results of \cite{NonZw05}, this model allows to
specify more precisely the dimension $d$ appearing in the scaling law. 
To our knowledge, so far the only
system for which the asymptotics corresponding to \eqref{e:Weyl-law} 
could be rigorously proven is the ``Walsh quantization'' of 
the symmetric open baker's map \cite{NonZw06}, which will be described 
in \S\ref{s:walsh}.

After counting resonances, the next step consists in studying the 
long-living resonant eigenstates $\varphi_j$ or $\tilde\varphi_j$.
Some results on this matter 
have been announced by M.Zworski and the first author in \cite{NonZw-gap}.
Interesting numerics were performed by M.~Lebental and coworkers for a 
model of open stadium billiard, relevant to describe an experimental micro-laser cavity
\cite{Leben06}. In the framework of quantum open maps, eigenstates of the open Chirikov
map have been numerically studied by Casati {\it et al.} \cite{CasMasShep99}; the authors
showed that, in the semiclassical
limit, the long-living eigenstates concentrate on the classical hyperbolic repeller. They
also found that the phase space structure of the eigenstates are very correlated with their 
decay rate. In this paper we will consider general ``quantizable'' open maps,
%mainly focus on a different map, namely the open baker, 
and formalize the above observations into rigorous statements. Our main result is
Theorem~\ref{thm:main} (see \S\ref{s:semiclass}), which 
shows that the {\it semiclassical measure} associated
to a sequence of quantum eigenstates is (up to subtleties due to discontinuities) 
necessary an {\em eigenmeasure} of the classical open map.
Such eigenmeasures are necessarily supported on the
the {\em backward trapped set}, which, in the case of a chaotic dynamics, 
is a fractal subset of the phase space: this motivated
the denomination of ``quantum fractal eigenstates'' used in \cite{CasMasShep99}.
Let us mention that eigenstates of quantized open maps have been studied 
in parallel by J.~Keating and coworkers \cite{KNPS06}.
Our theorem~\ref{thm:main} provides a rigorous version of
statements contained in their work.

Inspired by our experience with closed
chaotic systems, in \S\ref{s:abundance} we attempt to
classify
semiclassical measures among all possible eigenmeasures, in particular
for the open baker's map.
In the case of
the ``standard'' quantized open baker, the classification remains open.
In \S\ref{s:walsh} we consider a solvable model,
the Walsh quantization of the open baker's map, introduced in \cite{NonZw05,NonZw06}. 
For that model, one can explicitly construct some semiclassical measures and partially
answer the above questions. A further study of semiclassical measures for
the Walsh model will appear in a joint publication with J.~Keating, M.~Novaes and M.~Sieber.

\subsection*{Acknowledgments}
M.~Rubin thanks the Service de Physique Th\'eorique for hospitality 
in the spring 2005, during which this work was initiated. S.~Nonnenmacher
has been partially supported from the grant ANR-05-JCJC-0107-01
of the Agence Nationale de la Recherche.
Both authors are grateful to J.~Keating, M.~Novaes and M.~Sieber for
communicating their results concerning the Walsh-quantized baker before 
publication, and for interesting discussions. We also thank M.~Lebental
for sharing with us her preliminary results on the open stadium billiard,
and the anonymous referees for their stimulating comments.

%%%%%%%%%%%%%%%%%%%%%%%%%%%%%%%%%%%%%%%%%%%%%%%%%%%%%%%%%%%%%%%%%%%%%%%
%%%%%%%%%%%%%%%%%%%%%%%%%%%%%%%%%%%%%%%%%%%%%%%%%%%%%%%%%%%%%%%%%%%%%%%
\section{The open baker's map}
%%%%%%%%%%%%%%%%%%%%%%%%%%%%%%%%%%%%%%%%%%%%%%%%%%%%%%%%%%%%%%%%%%%%%%%
%%%%%%%%%%%%%%%%%%%%%%%%%%%%%%%%%%%%%%%%%
\subsection{Closed and open symplectic maps}\label{e:open-map} 
%%%%%%%%%%%%%%%%%%%%%%%%%%%%%%

Although many of the results we will present deal with a particular family of maps on the
2-torus, namely the family of baker's maps, we start by some general considerations on
open maps defined on a compact metric space $\hM$, equipped with a probability measure $\mu_L$. 
We borrow some ideas and notations from the 
recent review of Demers and Young \cite{DemYou06}. We start with a invertible map $\hT:\hM\to\hM$,
which we assume to be piecewise smooth, and to preserve the measure $\mu_L$ (when
$\hM$ is a symplectic manifold, $\mu_L$ is the Liouville measure).
We then {\em dig a hole} in $\hM$, that is a certain subset $H\subset \hM$, 
and decide that points falling in the hole are no more iterated, but rather
``disappear'' or ``go to infinity''. The hole is assumed to be a Borel subset of $\hM$.

Taking $M\defeq \hM\setminus H$, we are thus lead to consider 
the {\em open map} $T=\hT_{|M}:M\to \hM$, or equivalently, say that
$T$ sends points in the hole to infinity. By 
iterating $T$, we see that any point $x\in M$ has a certain time of escape $n(x)$, which is the
smallest integer such that $\hT^n(x)\in H$ (this time can be 
infinite). For each $n\in\IN^*$ we call 
\bequ
\label{e:M^n}
M^n=\set{x\in M,\ n(x)\geq n}=\bigcap_{j=0}^{n-1} \hT^{-j}(M)\,.
\eequ
This is the domain of definition of
the iterated map $T^n$. The {\em forward trapped set} for the open map $T$ is
made of the points which will never escape in the future:
\bequ\label{e:forward-TS}
\Gamma_- = \bigcap_{n\geq 1}M^n = \bigcap_{j= 0}^\infty \hT^{-j}(M)\,.
\eequ
These definitions allow us to split
the full phase space into a disjoint union
\bequ\label{e:split}
\hM=\big(\bigsqcup_{n=1}^\infty M_n\big)\sqcup \Gamma_-,
\eequ
where $M_1\defeq H$, and for each $n\geq 2$, $M_n\defeq M^{n-1}\setminus M^{n}$ 
is the set of points 
escaping exactly at time $n$.

We also consider the backward evolution given by the inverse map $\hT^{-1}$. 
The hole for this backwards 
map is the set $H^{-1}=\hT(H)$, and we call
$T^{-1}$ the restriction of $\hT^{-1}$ to $M^{-1}=\hT(M)$ 
(the ``backwards open map''). We also define $M^{-n}=\bigcap_{j=1}^n \hT^{j}(M)$, 
$M_{-1}=H^{-1}$, $M_{-n}=M^{-n+1}\setminus M^{-n}$ ($n\geq 2$). This leads to the
the {\em backward trapped set} 
\bequ\label{e:backward-TS}
\Gamma_+ = \bigcap_{j=1}^\infty \hT^{j}(M)\,,\quad\text{and the {\em trapped set} }
K=\Gamma_-\cap \Gamma_+\,.
\eequ
We also have a ``backward partition'' of the phase space:
\bequ\label{e:split2}
\hM=\big(\bigsqcup_{n=1}^\infty M_{-n}\big)\sqcup \Gamma_+.
\eequ
The dynamics of the open map $T$ is interesting if 
the trapped sets are not empty. This is be the case for the 
open baker's map we will study more explicitly (see \S\ref{s:baker-trapped}
and Fig.~\ref{f:Cantor}).

\medskip

In order to quantize the maps $\hT$ and $T$, one needs further assumptions.
In general, $\hM$ is
a {\em symplectic manifold}, $\mu_L$ its Liouville measure, and $\hT$ a canonical
transformation on $\hM$. Also, we will assume that $T$ is sufficiently regular (see
\S\ref{s:quantum}).

Yet, one may also consider ``quantizations'' on more general phase spaces, 
like in the axiomatic framework of Marklof and O'Keefe \cite{MarOK05}. 
As we will explain in \S\ref{s:Lipschitz}, 
the ``Walsh quantization'' of the baker's map is easier to analyze if we consider it
as the quantization of an open map on a certain symbolic space, which
is not a symplectic manifold. By extension, we also call ``Liouville'' the 
measure $\mu_L$ for this case.

%%%%%%%%%%%%%%%%%%%%%%%%%%%%%%%%%%%%%%%%%
\subsection{The open baker's map and its symbolic dynamics}\label{s:open-baker} 
%%%%%%%%%%%%%%%%%%%%%%%%%%%%%%
We present the closed and open maps which will be our central examples: the baker's
maps and their associated symbolic shifts.

\subsubsection{The closed baker}
The phase space of the baker's map is the 2-dimensional torus
$\hM=\t2\simeq [0,1)\times [0,1)$. A point on $\t2$ is described with
the coordinates $x=(q,p)$, which we call respectively position (horizontal) 
and momentum (vertical), to insist on the symplectic structure $dq\wedge dp$.
%%%%%%%%%%%%%%%
\begin{figure}[htbp]
\begin{center}
\includegraphics[width=13cm]{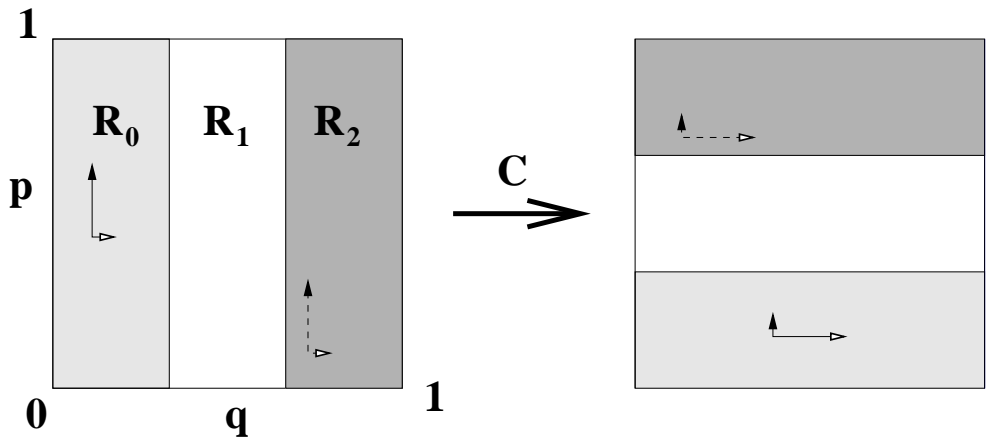}
\caption{\label{f:3-baker} Sketch of the closed baker's map $\hB_{\vr}$, and its
open counterpart $B_{\vr}$, for the case $\vr=\vr_{sym}=(1/3,1/3,1/3)$. 
The three rectangles form a Markov partition.}
\end{center}
\end{figure}
%%%%%%%%%%%%%%%
We split $\t2$ into three vertical rectangles $R_i$ with widths $(r_0,r_1,r_2)\defeq \vr$
(such that $r_0+r_1+r_2=1$, with all $r_\ep>0$), and first define a
closed baker's map on $\t2$ (see Fig.~\ref{f:3-baker}):
\bequ\label{e:closed-baker}
(q,p)\mapsto \hB_{\vr}(q,p)\defeq (q',p')= \left\{  \begin{array}{ll}
\big(\frac{q}{r_0},\,  p\,r_0\big) & \text{if}\  0 \leq q < r_0 \\
\big(\frac{q-r_0}{r_1}, \,  p\,r_1+r_0 \big)& \text{if}\  r_0 \leq q < r_0+r_1 \\
\big(\frac{q-r_0-r_1}{r_2}, \,  p\,r_2+r_1+r_0\big)& \text{if}\  1-r_2 \leq q < 1  \,.
\end{array}\right. 
\eequ
This map is invertible and symplectic on $\t2$. 
It is discontinuous on the boundaries of the rectangles $R_i$ but smooth (actually, affine)
inside them.
We now recall how $\hB_{\vr}$ can be conjugated with a symbolic dynamics. 
We introduce the right shift $\hat\sigma$ on the {\em symbolic space} 
$\Sigma=\set{0,1,2}^{\IZ}$:
$$
\bep=\ldots\ep_{-2}\ep_{-1}\cdot\ep_0\ep_1\ldots\in \Sigma\longmapsto 
\hat\sigma(\bep)=\ldots\ep_{-2}\ep_{-1}\ep_0\cdot\ep_1\ldots\,.
$$
The symbolic space $\Sigma$ can be mapped to the 2-torus as follows: every
bi-infinite sequence $\bep\in\Sigma$ is mapped to the point $x=J_{\vr}(\bep)$ with coordinates
\bequ\label{e:conjugation}
q(\bep)=\sum_{k=0}^\infty r_{\ep_0}r_{\ep_1}\cdots r_{\ep_{k-1}}\,\alpha_{\ep_k}\,,\qquad
p(\bep)=\sum_{k=1}^\infty r_{\ep_{-1}}r_{\ep_{-2}}\cdots r_{\ep_{-k+1}}\,\alpha_{\ep_{-k}}\,,
\eequ
where we have set $\alpha_0=0$, $\alpha_1=r_0$, $\alpha_2=r_0+r_1$. 
The position coordinate (unstable direction) depends on symbols 
on the right of the comma, while the momentum coordinate (stable direction)
depends on symbols on the left.
The map $J_{\vr}:\Sigma\to\t2$ is
surjective but not injective, for instance the sequences $\ldots \ep_{n-1}\ep_n 000\ldots$ 
(with $\ep_n\neq 0$) and $\ldots \ep_{n-1}(\ep_n-1) 222\ldots$ have the same image on $\t2$. 
For this reason, it is convenient to restrict $J_{\vr}$ to the subset $\Sigma'\subset\Sigma$
obtained by removing from $\Sigma$ the sequences ending by $\ldots 2222 \ldots$ on the
left or the right.
$\Sigma'$ is invariant through the shift $\hat\sigma$. 
The map $J_{\vr|\Sigma'}:\Sigma'\to\t2$ is now bijective, and it conjugates
$\hat\sigma_{|\Sigma'}$ with the baker's map $\hB_{\vr}$ on $\t2$:
\bequ\label{e:shift-A}
\hB_{\vr} = J_{\vr|\Sigma'}\circ \hat\sigma\circ (J_{\vr|\Sigma'})^{-1}\,.
\eequ
Any finite sequence 
$\bep=\ep_{-m}\ldots\ep_{-1}\cdot\ep_{0}\ldots\ep_{n-1}$ represents a {\em cylinder} 
$[\bep]\subset\Sigma$, which consists in the sequences sharing the same symbols 
between indices $-m$ and $n-1$. 
We call $J_{\vr}([\bep])$
a {\em rectangle} on the torus (sometimes we will note it $[\bep]$).
This rectangle has sides
parallel to the two axes; it has width $r_{\ep_0}r_{\ep_1}\cdots r_{\ep_{n-1}}$
and height $r_{\ep_{-1}}\cdots r_{\ep_{-m}}$.

For any triple, the Liouville measure on $\t2$ is the push-forward through $J_{\vr}$ 
of a certain Bernoulli measure on
$\Sigma$, namely $\mu^\Sigma_L= \nu^{\Sigma_-}_{\vr}\times \nu^{\Sigma_+}_{\vr}$
(see \S\ref{s:Bernoulli}).

%=======================================
\subsubsection{The open baker}\label{s:baker-trapped}
%=======================================

We choose to take for the hole the middle rectangle $H=R_1=\set{r_0\leq q<1-r_2}$.
We thus obtain an ``open baker's map'' $B_{\vr}$ defined on $M=\t2\setminus H$:
$$
(q,p)\mapsto B_{\vr}(q,p)\defeq (q',p')= \left\{  \begin{array}{ll}
\big(\frac{q}{r_0},\,  p\,r_0 \big) & \text{if}\  0 \leq q < r_0 \\
\quad\infty& \text{if}\  r_0 \leq q < 1-r_2 \\
\big(\frac{q-1+r_2}{r_2}, \,  p\,r_2+1-r_2\big)& \text{if}\  1-r_2 \leq q < 1 \,.
\end{array}\right. 
$$ 
The ``inverse map'' $B_{\vr}^{-1}$
is defined on the set $M^{-1}=B_{\vr}(M)$, that is outside the backwards hole 
$H^{-1}=B_{\vr}(H)=\set{r_0\leq p<1-r_2}$ (see Fig.~\ref{f:3-baker}, right).

Due to the choice of the hole, this open map is still easy to analyze 
through symbolic dynamics:
the hole $R_1$ is the image through $J_{\vr}$ of the set $\set{\ep_0=1}\cap \Sigma'$.
Let us define as follows the {\it open shift}
$\sigma$ on $\Sigma$:
\bequ\label{e:shift-B}
\bep\in \Sigma%=\ldots\ep_{-2}\ep_{-1}\cdot\ep_0\ep_1\ldots%\stackrel{B_{\vr}}{\longmapsto} 
\longmapsto \sigma(\bep)\defeq 
\begin{cases}\infty\quad\mbox{if}\quad \ep_0=1\\
\hat\sigma(\bep)%\ldots\ep_{-2}\ep_{-1}\ep_0\cdot\ep_1\ldots=\hat\sigma(\bep)
\quad \mbox{if}\quad\ep_0\in\set{0,2}\,.
\end{cases}
\eequ
$J_{\vr|\Sigma'}$ conjugates the open shift $\sigma_{|\Sigma'}$
with $B_{\vr}$ as in \eqref{e:shift-A}. Similarly, the backwards open shift 
$\sigma^{-1}$, which kills the sequences
s.t. $\ep_{-1}=1$ and otherwise moves the comma to the left, is conjugated with
$B_{\vr}^{-1}$.

\medskip

\begin{figure}[htbp]
\begin{center}
\includegraphics[width=16cm]{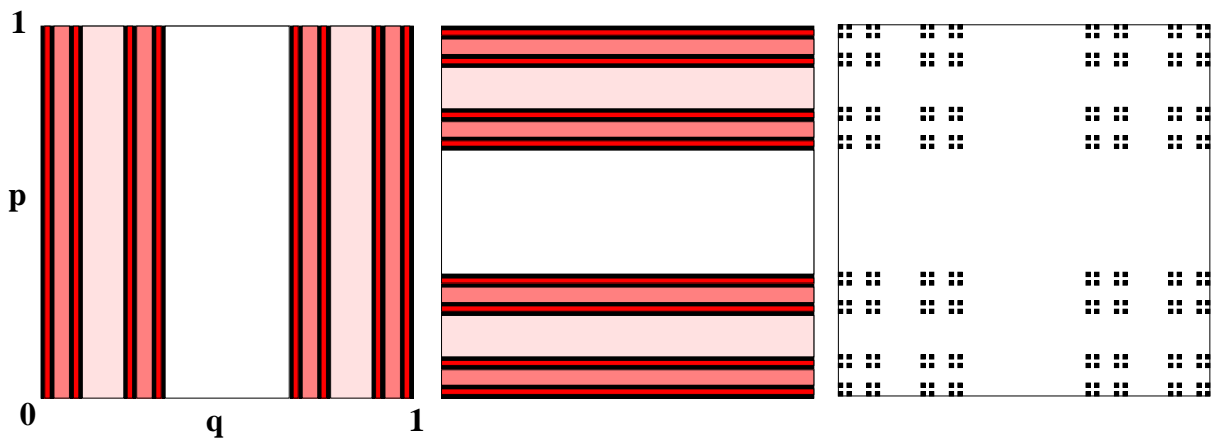}
\caption{\label{f:Cantor} In black we approximate the forward (left), 
backward (center)
and joint (right) trapped sets
for the symmetric open baker $B_{\vr_{sym}}$. 
On the left (resp. centre), red/gray scales (from white to
dark) correspond to points escaping 
at successive times in the future (resp. past), that is to sets $M_{n}$
(resp. $M_{-n}$), for $n=1,2,3,4$.}
\end{center}
\end{figure}

These conjugations allow to easily characterize the various trapped sets of $B_{\vr}$.
On the symbolic space $\Sigma$, the forward (resp. backward) trapped set of the
open shift $\sigma$ is given by the sequences $\bep$ such that 
$\ep_{n}\in\set{0,2}$ for all $n\geq 0$ (resp. for all $n<0$). To obtain the
trapped sets for the baker's map, we restrict ourselves on $\Sigma'$ and conjugate
by $J_{\vr|\Sigma'}$. The image sets are given by the direct products 
$\Gamma_-=\cC_{\vr}\times [0,1)$, $\Gamma_+=[0,1)\times \cC_{\vr}$ 
and $K=\cC_{\vr}\times\cC_{\vr}$,
where $\cC_{\vr}$ is (up to a countable set) 
the Cantor set on $[0,1)$ adapted to the partition $\vr$ 
(see Fig.~\ref{f:Cantor}).
%%%%%%%%%%%%%%%%%%%%
%%%%%%%%%%%%%%%%%

For future use, we define the subset $\Sigma''\subset\Sigma$ by
\bequ\label{e:Sigma''}
\Sigma''\defeq \Sigma\setminus
\big(\set{\ldots 222\cdot 0\ep_1\ep_2\ldots}\cup\set{\ldots 222\cdot 2\ep_1\ep_2\ldots}\cup 
\set{\ldots \ep_{-2}\ep_{-1}\cdot \ep_0 222\ldots}\big)\,.
\eequ
We notice that $\Sigma''\varsupsetneq\Sigma'$, and that $J_{\vr}(\Sigma\setminus \Sigma'')$ is
a subset of the discontinuity set of the map $B_{\vr}$ (see \S\ref{s:quantum-baker2}
and Fig.~\ref{f:gamma+}). 
The map $J_{\vr}$ realizes a kind of {\it semiconjugacy} between $\sigma_{|\Sigma''}$ and $B$:
\bequ\label{e:semiconj}
\forall \bep\in \Sigma'',\qquad  J_{\vr}\circ\sigma(\bep)=B_{\vr}\circ J_{\vr}(\bep).
\eequ
Above it is understood that both sides are ``sent to infinity'' if $\ep_{0}=1$. 

%%%%%%%%%%%%%%%%%%%%%%%%%%%%%%%%%%%%%%%%%
\subsection{Eigenmeasures of open maps}\label{s:eigenmeasures0} 
%%%%%%%%%%%%%%%%%%%%%%%%%%%%%%%%%%%%%%%%%

Before defining eigenmeasures of
open maps, we briefly recall how invariant measures emerge in the study of the
quantized {\it closed} maps.

\subsubsection{Quantum ergodicity for closed chaotic systems}
\label{s:eigenstates-IM} 

The quantum-classical correspondence between a closed symplectic map
$\hT$ and its quantization $\hT_{N}$ (see \S\ref{s:quantum}) 
has one important consequence:
in the semiclassical limit $N\to\infty$, stationary states of the quantum system (that is,
eigenstates of $\hT_{N}$) should reflect the stationary properties of the
classical map, namely its invariant measures. 
To be more precise, 
to any sequence of eigenstates $(\psi_N)_{N\to \infty}$ of the quantum map, 
one can associate at least one {\bf semiclassical measure} (see \S\ref{s:semicl-measure}).
Omitting problems due to the discontinuities of $\hT$, the quantum-classical
correspondence implies that a semiclassical measure $\mu$ must be 
invariant w.r.to $\hT$:
\bequ\label{e:invariant}
\hT^*\,\mu=\mu\quad\Longleftrightarrow 
\quad\text{for any Borel set } S\subset \t2,\quad \mu(\hT^{-1}(S))=\mu(S)\,.
\eequ
$\mu$ is then also invariant w.r.to the inverse map $\hT^{-1}$.

If the map $\hT$ is ergodic w.r.to the Liouville measure $\mu_{L}$ on $\hM$, 
the quantum ergodicity theorem (or Schnirelman's theorem \cite{Schni74})
states that, for ``almost any'' sequence $(\psi_N)_{N\to \infty}$,
there is a unique associated semiclassical measure, which is $\mu_{L}$ itself.

Such a theorem was first proven for eigenstates of the Laplacian on compact
Riemannian manifolds with ergodic geodesic flow \cite{Zel87,CdV85}, then for more general 
Hamiltonians \cite{HelMarRob87}, billiards \cite{GerLei93,ZelZwo96}
and maps \cite{BouzDB96,Zel96}. Quantum ergodicity for piecewise smooth
maps was proven in a general setting in \cite{MarOK05}, and the particular case of the
baker's map was treated in \cite{DENW06}. Finally, in \cite{AN06} it was shown that
the (closed) Walsh-baker's map, seen as the quantization of the right shift $\hat\sigma$
on $(\Sigma,\mu_L)$, also satisfies quantum ergodicity with respect to $\mu_L$.

It is generally unknown whether there exist ``exceptional sequences'' of eigenstates, 
converging to a different invariant measure. 
The absence of such sequences is expressed by the
quantum unique ergodicity conjecture \cite{RudSar94}, which has been proven only for
systems with arithmetic properties \cite{Linden06,KurRud00}. This conjecture has 
been disproved for some specific systems enjoying
large spectral degeneracies at the quantum level, allowing for
sufficient freedom to build up partially localized eigenstates \cite{FNdB03,AN06,Kelmer05}. 
Some special eigenstates of the standard quantum baker with interesting multifractal
properties have been numerically identified \cite{MeenakLaksh04}, 
but their persistence in the semiclassical limit remains unclear.

%%%%%%%%%%%%%%%%%%%%%%%%%%%%%%%%%%%%%%%%%
\subsubsection{Eigenmeasures of open maps}\label{s:eigenmeasures} 
%%%%%%%%%%%%%%%%%%%%%%%%%%%%%%%%%%%%%%%%%

We now dig the hole $H=\hM\setminus M$, and consider the open map $T=\hT_{|M}$. 
Eigenmeasures (also called conditionally invariant measures) of maps ``with holes'' have been less
studied than their invariant counterparts. The recent article of Demers and Young \cite{DemYou06}
summarizes most of the properties of these measures, for maps enjoying various dynamical
properties. Below we describe some of these properties for general maps, before 
being more specific in the
case of the baker's map.

A probability measure $\mu$ on $\hM$ which is invariant
through $T$ {\em up to a multiplicative factor} will be called
an eigenmeasure of $T$:
\bequ\label{e:eigenmeasure}
T^*\,\mu=\Lambda_\mu\,\mu
\quad\Longleftrightarrow\quad 
\text{for any Borel set } S\subset\hM,\quad \mu(T^{-1}(S))=\Lambda_\mu\,\mu(S)
\,. 
\eequ
Here $\Lambda_\mu\in [0,1]$ (or rather $\gamma_\mu=-\log\Lambda_\mu$) 
is called the ``escape rate'' or ``decay rate'' of the eigenmeasure $\mu$, and is
given by $\Lambda_{\mu}=\mu(\cS)$. It corresponds
to the fact that a fraction of the particles in the support of $\mu$ 
escape at each step. Our definition slightly differs from the one in \cite{DemYou06}:
their measures are supported on $M$ and normalized there, while we choose the normalization
$\mu(\hM)=1$.

Here are some simple properties:
%%%%%%%%%%%%%%%
\begin{prop}
Let $\mu$ be an eigenmeasure of $T$ with decay rate $\Lambda_\mu$. \\
If $\Lambda_\mu=0$, $\mu$ is supported in the hole $H$. \\
If $\Lambda_\mu=1$, $\mu$ is supported in the trapped set $K$ (invariant measure).\\
If $0<\Lambda_\mu<1$, $\mu$ is supported on the set $\Gamma_+\setminus K$.
\end{prop}
%%%%%%%%%%%%%%%

%%%%%%%%%%%%%%%
\begin{figure}[htbp]
\begin{center}
\includegraphics[width=8cm]{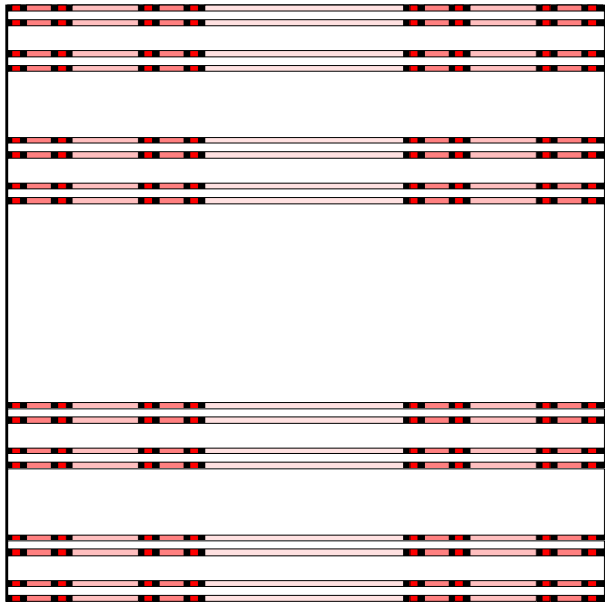}
\caption{\label{f:Cantor6} Various components $\Gamma_+^{(n)}$
of the forward trapped set for the open baker $B_{\vr_{sym}}$. 
Various red/grey scales (from light to dark) correspond to $n=1,2,3,4$.}
\end{center}
\end{figure}
%%%%%%%%%%%%%%%
Following the decomposition \eqref{e:split}, the set $\Gamma_+\setminus K$ can be split into a 
disjoint union (see Fig.~\ref{f:Cantor6}):
$$
\Gamma_+\setminus K=\bigsqcup_{n\geq 1} \Gamma_+^{(n)},\quad\text{where}\qquad
\Gamma_+^{(n)}\defeq \Gamma_+\cap M_{n}=
T^{-n+1}\Gamma_+^{(1)},
\quad n\geq 1\,.
$$
Following \cite[Thm. 3.1]{DemYou06}, any
eigenmeasure can be constructed as follows.
%&&&&&&&&&&&&&&&&&&&&&&&&&&&&&&&&&&&&
\begin{prop}\label{p:eigen-decompo}
Take some $\Lambda\in [0,1)$ and $\nu$ an arbitrary Borel probability
measure on $\Gamma_+^{(1)}=\Gamma_+\cap H$.
Define the probability measure $\mu$ on $\t2$ as follows:
\bequ\label{e:eigen-decompo}
\mu= %(1-\Lambda) \sum_{n\geq 0}\Lambda^n\,(\hT^*)^n\,\nu=
(1-\Lambda) \sum_{n\geq 0}\Lambda^n\,(T^*)^n\,\nu\,.
\eequ
Then $T^*\,\mu=\Lambda\,\mu$.
All $\Lambda$-eigenmeasures of $T$ can be written this way.
\end{prop}
%&&&&&&&&&&&&&&&&&&&&&&&&&&&&&&&&&&&&
This construction shows that, as long as $\Gamma_+^{(1)}$ is not empty 
(that is, there exists at least a point $x_0$ trapped in the past but escaping
in the future), there are plenty of eigenmeasures. 

The case where the map $\hT$ is uniformly hyperbolic has been studied in detail
by Chernov and Markarian \cite{Mark96,CherMar97}. The set $\Gamma_+^{(1)}$ is then 
uncoutable, and it is foliated by the unstable foliation. These authors 
focussed on eigenmeasures of the open map $T$ which are
absolutely continuous along the unstable direction. 
Even with this condition, one has plenty of eigenmeasures 
(we remind that
for a closed hyperbolic map, unstable absolute continuity is satisfied for a 
unique invariant measure, namely the SRB measure).

%=======================================
\subsubsection{Pure point eigenmeasures}\label{s:pp} 
%=======================================

Applying the recipe of Proposition~\ref{p:eigen-decompo} to the Dirac measure $\delta_{x_0}$ on
an arbitrary point $x_0\in\Gamma_+^{(1)}$, we obtain a simple, pure point $\Lambda$-eigenmeasure
supported on the backwards trajectory $\big(x_{-n}=T^{-n}(x_0)\big)_{n\geq 0}$. 
For any $\Lambda\in (0,1)$, we call this eigenmeasure
\bequ\label{e:mu-traject}
\mu_{x_0,\Lambda}\defeq (1-\Lambda)\sum_{n\geq 0} \Lambda^n\,\delta_{x_{-n}}\,.
\eequ
We recall that the pure point {\em invariant} 
measures of $T$ are localized on periodic orbits of $T$
on $K$ (which, for the case of a horseshoe map $T_{|K}$, form a countable family). 
On the opposite, 
for each $\Lambda\in [0,1)$ the family of pure point eigenmeasures 
$\mu_{x_0,\Lambda}$ is labelled by all $x_0\in \Gamma_+^{(1)}$ (an uncountable set).

%=======================================
\subsubsection{Natural eigenmeasure}\label{s:natural} 
%=======================================

As discussed in \cite{DemYou06}, one may search for the definition of a {\em natural} eigenmeasure of
$T$. The simplest (``ideal'') definition reads as follows: for any 
initial measure $\rho$ absolutely continuous w.r.to $\mu_L$, and
such that $\rho(\Gamma_-)> 0$,
\bequ\label{e:SRB-open}
\mu_{nat}=\lim_{n\to\infty} \frac{T^{*n}\,\rho}{\|T^{*n}\,\rho\|}\,,\qquad\text{where}\ 
\|\mu\|\defi \mu(\hM)\,.
\eequ
It was shown in \cite{Mark96,CherMar97}
that, for a hyperbolic open map, the limit exists and is independent of $\rho$. Besides,
the natural measure is then absolutely continuous along the unstable direction.
Yet, for a general open map the above limit does not necessarily exist, 
or it may depend on the initial distribution $\rho$ \cite{DemYou06}.

In case \eqref{e:SRB-open} holds, 
the eigenvalue $\Lambda_{nat}=\mu_{nat}(M)$ associated with this measure is generally called
``the decay rate of the system'' by physicists.
For a {\em closed} chaotic map $\hT$, 
the natural measure is the Liouville 
measure $\mu_{L}$ (indeed, $\hT^{*n}\rho\to \mu_{L}$ is equivalent with
the fact that $\hT$ is {\em mixing} with respect to $\mu_{L}$). 
As explained in \S\ref{s:eigenstates-IM}, the quantum ergodicity theorem 
shows that this particular invariant measure is ``favored'' by quantum mechanics.
One interesting question we will address is the relevance of
$\mu_{nat}$ with respect to the quantized open map $T_N$.

%%%%%%%%%%%%%%%%%%%%%%%%%%%%%%%%%%%
\subsubsection{Bernoulli eigenmeasures of the open baker}\label{s:Bernoulli} 
%%%%%%%%%%%%%%%%%%%%%%%%%%%%%%%%%%%

We now focus on the open baker $B_{\vr}$ and the open shift $\sigma$ it is conjugated with,
and construct a family of eigenmeasures, called Bernoulli eigenmeasures. Some of
these measures will appear in \S\ref{s:walsh} as
semiclassical measures for the Walsh-quantized open baker.

The first equality in \eqref{e:conjugation} maps
the set $\Sigma_+$ of one-sided sequences $\cdot \ep_0\ep_1\ldots$ to
the position interval $[0,1]$. By a slight abuse, we also call this
mapping $J_{\vr}$. We will first construct Bernoulli measures on the
symbol spaces $\Sigma_+$ and $\Sigma$, and then push them on 
the interval or the torus using $J_{\vr}$. 

Let us recall the definition of Bernoulli measures on $\Sigma_+$. Choose
a {\em weight distribution} $\vP=(P_0,P_1,P_2)$, 
where $P_\ep\in[0,1]$ and $P_0+P_1+P_2=1$. We define the measure $\nu^{\Sigma_+}_{\vP}$
on $\Sigma_+$ as follows: 
for any $n$-sequence $\bep=\cdot\ep_0\ep_1\cdots\ep_{n-1}$, the weight of the cylinder $[\bep]$
is given by 
$$
\nu^{\Sigma_+}_{\vP}([\bep])=P_{\ep_{0}}\cdots P_{\ep_{n-1}}\,.
$$
The push-forward on $[0,1]$ of this measure will also be called a Bernoulli measure,
and called $\nu_{\vr,\vP}=J_{\vr}^*\,\nu^{\Sigma_+}_{\vP}$.
If we take $P_\ep=r_\ep$ for $\ep=0,1,2$, we recover $\nu_{\vr,\vr}=\nu_{Leb}$ the Lebesgue 
measure on $[0,1]$. 
If for some $\ep\in \set{0,1,2}$ we take $P_\ep=1$, we get for $\nu_{\vr,\vP}$ 
the Dirac measure at the point $q(\cdot \ep\ep\ep\ldots)$, which takes the respective
values $0$, $\frac{r_0}{1-r_1}$ and $1$. 
For any other distribution $\vP$,
the Bernoulli measure $\nu_{\vr,\vP}$ is purely singular
continuous w.r.to the Lebesgue measure. Fractal properties of the measures 
$\nu_{\vr,\vP}$ were studied in \cite{HJPPS86}. 
 If the weight $P_\ep$ vanishes for a single $\ep$ (e.g. $P_1=0$),
$\nu_{\vr,\vP}$ is supported on a Cantor set (e.g. $\cC_{\vr}$).

A Bernoulli measure on $\nu^{\Sigma_-}_{\vP}$ can be defined similarly.
By taking products of two Bernoulli measures, one easily constructs eigenmeasures of 
the open shift $\sigma$ or the open baker $B_{\vr}$.
%%%%%%%%%%%%%%%%%%%
\begin{prop}
Take any weight distribution $\vP$ such that $P_1<1$. Then the following hold.

$i)$ there is a unique auxiliary distribution,
namely $\vP^*=\set{\frac{P_0}{P_0+P_2},0,\frac{P_2}{P_0+P_2}}$,
such that the product measure
$$
\mu^\Sigma_{\vP}\defeq \nu^{\Sigma_+}_{\vP}\times\nu^{\Sigma_-}_{\vP^*}
$$
is an eigenmeasure of the open shift $\sigma$. The corresponding decay rate is
\bequ
\Lambda_{\vP}=1-P_1=P_0+P_2\,.
\eequ
$ii)$ for any $\vr$, the push-forward 
$\mu_{\vr,\vP}=J_{\vr}^*\,\mu^\Sigma_{\vP}$
is an eigenmeasure of the open baker $B_{\vr}$.
\end{prop}
%%%%%%%%%%%%%%%%%%%
By definition, the product measure has the following weight on a cylinder 
$[\bep]=[\ep_{-m}\ldots\ep_{n-1}]$:
$$
\mu^\Sigma_{\vP}([\bep])=P^*_{\ep_{-m}}\ldots P^*_{\ep_{-1}}\,P_{\ep_{0}}\cdots P_{\ep_{n-1}}\,.
$$
The proof of the first statement is straightforward.
There is a slight subtlety concerning push-forwards of $\sigma$-eigenmeasures, 
which is resolved in the following 
%============
\begin{prop}\label{l:eigenmeasure-push}
Let $\mu^\Sigma$ be an eigenmeasure of the open shift $\sigma:\Sigma\to\Sigma$. 
If $\mu^\Sigma$ does not charge the subset $\Sigma\setminus\Sigma''$ described in \eqref{e:Sigma''}, 
that is if $\mu^\Sigma(\Sigma\setminus\Sigma'')=0$, then its push-forward
$\mu=J_{\vr}^*\mu^{\Sigma}$ on $\t2$ is an eigenmeasure of $B_{\vr}$.
\end{prop}
%============
\begin{proof}
Take any Borel set $S\in\t2$.
Then the following identities hold:
$$
\mu(B_{\vr}^{-1}(S))\defeq \mu^\Sigma(J_{\vr}^{-1}\circ B_{\vr}^{-1}(S))
=\mu^\Sigma(\sigma^{-1}\circ J_{\vr}^{-1}(S))%\\
=\Lambda \, \mu^\Sigma(J_{\vr}^{-1}(S))
\defeq \Lambda \, \mu(S)\,.
$$
The second equality is a consequence of the semiconjugacy \eqref{e:semiconj}, which implies
the fact that the symmetric difference of the sets $J_{\vr}^{-1}\circ B_{\vr}^{-1}(S)$ and 
$\sigma^{-1}\circ J_{\vr}^{-1}(S)$
is necessarily a subset of $\Sigma\setminus\Sigma''$.
\end{proof}
The condition $\mu^\Sigma(\Sigma\setminus\Sigma'')=0$ in the proposition cannot be removed. 
Indeed, by applying the 
construction of proposition~\ref{p:eigen-decompo} to an initial ``seed'' 
supported on $\set{\ldots 222\cdot 1\ep_1\ldots}$, one obtains an eigenmeasure of $\sigma$ 
charging $\Sigma''$, and such that its push-forward is not an eigenmeasure of $B_{\vr}$.

To prove the second point of the Proposition, we remark that 
the only Bernoulli eigenmeasure $\mu^\Sigma_{\vP}$ charging $\Sigma\setminus\Sigma''$
is the Dirac measure on the sequence $\ldots 2222\ldots$, which is
pushed-forward to the delta measure at the origin of $\t2$. 
$\hfill\square$

If we take $\vP=\vr$, the push-forward is
the natural eigenmeasure $\mu_{\vr,\vr}=\mu_{nat}$ of the open baker $B_{\vr}$.
If $\vP\neq\vP'$, the Bernoulli eigenmeasures $\mu^\Sigma_{\vP}$ and
$\mu^\Sigma_{\vP'}$
are {\em mutually singular} (there exists disjoint Borel subsets $A$, $A'$ of $\Sigma$
such that $\nu_{\vr,\vP}(A)=\nu_{\vr,\vP'}(A')=1$),
eventhough they may share the same decay rate.
Except in the case $\vP=(1,0,0)$, $\vP'=(0,0,1)$, the push-forwards 
$\mu_{\vr,\vP}$ and $\mu_{\vr,\vP'}$ are also mutually singular.

%%%%%%%%%%%%%%%%%%%%%%%%%%%%%%%%%%%%%%%%%
%%%%%%%%%%%%%%%%%%%%%%%%%%%%%%%%%%%%%%%%%
\section{Quantized open maps}
%%%%%%%%%%%%%%%%%%%%%%%%%%%%%%%%%%%%%%%%%
%%%%%%%%%%%%%%%%%%%%%%%%%%%%%%%%%%%%%%%%%

%%%%%%%%%%%%%%%%%%%%%%%%%%%%%%%%%%%%%%%%%
\subsection{``Axioms'' of quantization}\label{s:quantum}
%%%%%%%%%%%%%%%%%%%%%%%%%%%%%%%%%%%%%%%%%
For appropriate $2D$-dimensional compact symplectic manifolds $\hM$, one may
define a sequence of ``quantum'' Hilbert spaces $(\hn)_{N\to\infty}$ of 
finite dimensions $N$, which are related with Planck's constant by
$N\sim \hbar^{-D}$. 
Quantum states are normalized vectors in $\hn$. One
also wants to quantize observables, that is functions $a\in C^\infty(\hM)$,
into operators $\Op_N(a)$ on $\hn$. An invertible (resp. open) map $\hT$ (resp. $T$)
is quantized into a family of
a {\em unitary} (resp. contracting) operators $\hT_N$ (resp. $T_N$), 
which satisfies certain properties when $N\to\infty$ (see below). 

For the example we will treat explicitly (the open baker's map), 
the propagators of the closed and open maps are related by
$T_N=\hT_N\circ\Pi_{M,N}$,
where $\Pi_{M,N}$ is a projector associated with the subset $M$: it kills 
the quantum states microlocalized in the hole, while keeping unchanged
the states microlocalized inside $M$ \cite{SaVa96}. 
Yet, in general the propagator
$T_N$ can be defined without having to construct $\hT_N$ beforehand.

The proof of our main result, Theorem~\ref{thm:main}, only uses some
``minimal'' properties of the quantized observables and
maps. These properties were presented as
``quantization axioms'' by Marklof and O'Keefe in the case of closed maps \cite{MarOK05}.
We adopt the same approach, namely define quantization through these ``minimal axioms'',
and state our result in this general framework.
Afterwards, we will check that these
axioms are satisfied for the quantized open baker.

%================================
\subsubsection{Axioms on observables}\label{s:axioms-obs}
%================================

All axioms will describe properties of the quantum operators in the semiclassical limit 
$N\to\infty$. With a slight abuse of notation, we 
write $A_N\sim B_N$ when two families of operators $(A_N)$, $(B_N)$ on $\hn$ satisfy
$$
\norm{A_N-B_N}_{\cL(\hn)}\Nto8 0\,.
$$
The axioms concerning the quantization of observables
read as follows \cite[Axiom~2.1]{MarOK05}. 
\begin{Def}\label{d:axiom-obs}
For any $a\in C^\infty(\hM)$, the operators
$\Op_N(a)$ on $\hn$ must satisfy, in the limit $N\to\infty$:
\bequ
\begin{split}\label{e:axiom-obs}
\Op_N(\bar{a})\sim \Op_N(a)^{\dagger}\,&\qquad \text{(asymptotic hermiticity)}\\
\Op_N(a)\,\Op_N(b)\sim \Op_N(ab)\,&\qquad \text{(0-th order symbolic calculus)}\\
\lim_{N\to\infty} N^{-1}\,\Tr \Op_N(a)=\int_{\hM}a\,d\mu_{L}&\qquad \text{(normalization)}\,.
\end{split}
\eequ
\end{Def}
These axioms are satisfied by all standard quantization recipes (e.g. geometric or
Toeplitz quantization on K\"ahler manifolds \cite{Zel96}). Notice that they do not
involve the symplectic structure on $\hM$, and can thus be extended to more general phase
spaces.

%================================
\subsubsection{Semiclassical mesures}\label{s:semicl-measure}
%================================
From there, we may define semiclassical measures associated with sequences of (normalized)
quantum states $(\psi_N\in\hn)_{N\to\infty}$. Such a sequence is said to converge to the
distribution $\mu$ on $\hM$ iff, for any observable $a\in C^\infty(\hM)$, 
\bequ\label{e:semiclass-measure}
\la \psi_N,\,\Op_N(a)\psi_N\ra\Nto8 \int_{\hM}a\,d\mu\,.
\eequ
From the above axioms, one can show that $\mu$ is necessarily 
a probability measure on $\hM$,
which is called the semiclassical measure associated with $(\psi_N\in\hn)$.
By weak compactness, from any sequence $(\psi_N\in\hn)$ one can always extract a
subsequence $(\psi_{N_k})$ converging in the above sense to a certain measure. The latter
is called {\it a} semiclassical measure of the sequence $(\psi_N)$.

%==============================
\subsubsection{Axioms on maps}\label{s:axiom-maps}
%==============================

In the axiomatic framework of \cite{MarOK05}, the conditions satisfied by 
the unitary propagators $(\hT_N)$ quantizing a closed map $\hT$ 
consist in some form of quantum-classical correspondence (Egorov's theorem), 
when evolving quantum observables $\Op_N(a)$ through these propagators. 
However, if $\hT$ is discontinuous (or nonsmooth) at some
points, its propagator $\hT$ will exhibit diffraction phenomena 
around these ``singular'' points,
which alter the propagation properties. At the classical level,
a smooth observable $a\in C^\infty(\hM)$ is transformed into 
a smooth observable $a\circ \hT\in C^\infty(\hM)$ only if $a$ vanishes
near the singular points of $\hT^{-1}$.
As a result, the quantum-classical correspondence
is a reasonable axiom only if the observable $a$ is
supported ``far away'' from the singular set of $\hT^{-1}$..

We now specify our assumptions on the open map $T:M\to\hM$. Firstly, the
hole $H$ will be a ``nice'' set, that is a set with nonempty interior and such 
that $\partial H$ has Minkowski content zero 
($i.e.$ the volume of its $\eps$-neighbourhood vanishes when $\eps\to 0$).
We also assume that $M$, the domain of definition of $T$, can be decomposed using 
finitely or countably many open connected sets $O_i$:
$\overline{M}=\overline{\cup_{i\geq 1}O_i}$, with $O_i\cap O_j=\emptyset$, and for each $i$, 
$T_{|O_i}:O_i\to T(O_i)$ is a smooth canonical diffeomorphism, 
with all derivatives uniformly bounded. 
Let us split the hole into
$H=\ring{H}\sqcup DH$.
The {\it continuity set} (resp. {\it discontinuity set}) of the map $T$ is defined as
$$
C(T)=\sqcup_{i\geq 1}O_i\subset \ring{M}\,,
\quad\text{resp.}\quad D(T)\defi (M\setminus C(T))\sqcup DH\,.
$$
We assume that $D(T)$ has Minkowski content zero. We have the decomposition
$\hM=C(T)\sqcup \ring{H}\sqcup D(T)$. A similar decomposition holds for the inverse map:
\bequ\label{e:inverse-decompo}
\hM=C(T^{-1})\sqcup \ring{H}^{-1}\sqcup D(T^{-1}),\quad\text{with}\quad 
C(T^{-1})=T\big(C(T)\big)\,.
\eequ
Adapting the axioms of \cite{MarOK05} to the case of open maps, we set as follows
the characteristic
property of the operators $(T_N)_{N\to\infty}$ quantizing $T$.
\begin{Def}\label{d:axiom}
We say that the operators $\big(T_N\in\cL(\hn)\big)_{N\to\infty}$ quantize
the open map $T$ iff
\begin{itemize}
\item for $N$ large enough, $\norm{T_N}_{\cL(\hn)}\leq 1$ 
\item for any observable $a\in C^\infty_c\big(C(T^{-1})\sqcup \ring{H}^{-1}\big)$, we
have in the limit $N\to\infty$
\bequ\label{e:axiom2-map}
T_N^{\dagger}\,\Op_N(a)\,T_N \sim \Op_N\big(\bbbone_{M}\times(a\circ T)\big)\,.
\eequ
\end{itemize}
Here $C^\infty_c(S)$ indicates the smooth functions compactly supported inside $S$,
and $\bbbone_{M}$ is the characteristic function on $M$.
\end{Def}
Notice that, if $a$ is supported inside $C(T^{-1})\sqcup \ring{H}^{-1}$, 
the function $\bbbone_{M}\times(a\circ T)$ is well-defined,
smooth and supported inside $C(T)$. 
The factor $\bbbone_{M}$ ensures that
an observable supported inside $H^{-1}$ is ``semiclassically killed'' 
by the evolution through $T_N$.
The condition \eqref{e:axiom2-map} reminds of the definition of a
``quantized weighted relation'' introduced in \cite{NonZw06}, but it is less precise
(it only describes the lowest order in $\hbar$).

\begin{rem}
Going back to the problem of potential scattering mentioned in the introduction,
we expect the operators $T_N$ to share some spectral
properties with the propagator $\exp\big(-\i H_{W,\hbar}/\hbar\big)$ of the
``absorbing Hamiltonian'', in the semiclassical limit.
The eigenvalues $\set{\lambda_{j}}$ of $T_{N}$ 
should be compared with $\set{\e^{-\i \tilde{z}_j/\hbar}}$, or with 
$\set{\e^{-\i z_j/\hbar}}$, where $\set{\tilde{z}_j}$ (resp. $\set{z_j}$) are the
eigenvalues of $H_{W,\hbar}$ (resp. the resonances of $H_\hbar$). Similarly, 
the eigenstates of $T_N$ should share some microlocal properties with the eigenfunctions
$\tilde\varphi_j$ of $H_{W,\hbar}$ (resp. the resonant states $\varphi_j$ of $H_\hbar$)
inside the interaction region. Accordingly, we will sometimes 
call ``resonances'' and ``resonant
eigenstates'' the eigenvalues/states of quantized open maps.
\end{rem}

%%%%%%%%%%%%%%%%%%%%%%%%%%%%%%%%%%%%%%%%%
\subsection{The quantum open baker's map}\label{s:quantum-baker}
%%%%%%%%%%%%%%%%%%%%%%%%%%%%%%%%%%%%%%%%%

For any dimension $N=(2\pi\hbar)^{-1}$, the quantum Hilbert space $\hn$ 
adapted to the torus phase space is spanned by the
orthonormal {\em position basis} $\set{\bq_j,\ j=0,\ldots,N-1}$, localized at the 
discrete positions $q_j=\frac{j}N$. 

There are several standard ways to quantize observables on $\t2$: Weyl quantization,
Toeplitz (or anti-Wick) quantizations \cite{BouzDB96}, or Walsh quantization \cite{AN06}. 
Weyl and anti-Wick quantizations are
equivalent with each other in the semiclassical limit, in the sense that 
$\Op_N^{W}(a)\sim  \Op_N^{AW}(a)$ for any smooth observable. 
On the opposite, the
Walsh quantization (see \S\ref{s:walsh0}) is not equivalent 
to the previous ones.

After recalling the definition of the anti-Wick quantization on $\t2$ and
the ``standard'' quantization of the baker's map,
following the original approach of Balazs-Voros, Saraceno, 
Saraceno-Vallejos \cite{BV89,Sar90,SaVa96}, we check that the latter
satisfies Axioms~\ref{d:axiom}
with respect to the anti-Wick quantization.

%=====================================================
\subsubsection{Anti-Wick quantization of observables}\label{s:Husimi} 
%=====================================================
We recall the definition and properties of coherent states on $\t2$,
which we use to construct the anti-Wick quantization of observables,
and by duality the Husimi representation
of quantum states \cite{BouzDB96}.
The Gaussian coherent state in $L^2(\IR)$, localized at the phase space
point $\bx=(q_0,p_0)\in\IR^2$ and with squeezing parameter $\ssigma>0$, 
is defined by the normalized wavefunction
\begin{equation}
  \label{e:planeCS}
  \Psi_{\bx,\ssigma}(q)\defeq  \left( \frac{\ssigma}{\pi\hbar}\right)^{1/4}
   \,\e^{-\i \frac{p_0 q_0}{2\hbar}}\,
\e^{\i \frac{p_0 q}{\hbar}}\,
\e^{-\ssigma \frac{(q-q_0)^2}{2\hbar}}\,.
\end{equation}
When $\hbar=(2\pi N)^{-1}$, 
that state can be periodized on the torus, to yield the torus coherent state
$\psi_{\bx,\ssigma}\in\hn$ with following components in the basis $\set{\bq_j}$:
\bequ\label{e:torusCS}
\la\bq_j, \psi_{\bx,\ssigma}\ra=\frac{1}{\sqrt{N}}\,\sum_{\nu\in\IZ}
\Psi_{\bx,\ssigma}(j/N+\nu),\qquad
j=0,\ldots,N-1\,.
\eequ
For $s>0$ fixed, these states are asymptotically normalized when $N\to\infty$.

To any squeezing $\ssigma>0$ and inverse 
Planck's constant $N\in\IN$ we
associate the anti-Wick (or Toeplitz) quantization
\bequ\label{e:anti-Wick}
f\in C^\infty(\t2)\longmapsto
\OpAW(f)\defi \int_{\t2}  |\psi_{\bx,\ssigma}\ra\la\psi_{\bx,\ssigma}|\;f(\bx)\;N\,d\bx\,,
\eequ
which satisfies the Axioms~\eqref{d:axiom-obs} \cite{BouzDB96}.
By duality, this quantization defines, for any state $\psi\in\hn$, 
a Husimi distribution $H^\ssigma_\psi$:
$$
\forall f\in C^\infty(\t2),\qquad H^\ssigma_\psi(f)\defeq \la\psi,\OpAW(f)\,\psi\ra\,.
$$
For $\norm{\psi}=1$, this distribution is a probability measure, with density given by 
the (smooth, nonnegative) Husimi function 
\bequ\label{e:Husimi}
H^\ssigma_{\psi}(\bx)= N\,|\la \psi_{\bx,\ssigma},\psi\ra|^2\,,\quad
\bx\in\t2\,.
\eequ
Applying the definition \eqref{e:semiclass-measure} to the present framework, 
a sequence of states $(\psi_N\in\hn)_{N\to\infty}$ converges
to the measure $\mu$ on $\t2$ iff, for any given $\ssigma>0$, the Husimi measures
$(H^{\ssigma}_{\psi_N})_{N\to\infty}$ weak-$*$ converge to the measure $\mu$.

Following Schubert's work \cite{Roman-PhD}, 
one can also consider anti-Wick 
quantizations  (and dual Husimi measures)
in which the squeezing parameter $\ssigma$ in the integral \eqref{e:anti-Wick} 
depends on the phase space point.
Adapting the proofs of \cite{Roman-PhD} to the torus setting, one
shows that all these quantizations are equivalent to one another:
%+++++++++++++++++++++++++++++
\begin{prop}\label{p:quant}
Choose two functions  $\ssigma_1,\ssigma_2\in C^\infty(\t2,(0,\infty))$.
Then the two associated anti-Wick quantizations become close
to one another when $N\to\infty$:
$$
\forall f\in C^\infty(\t2),\qquad \norm{{\rm Op}^{{\rm AW}\!,\ssigma_1}_N(f)-
{\rm Op}^{{\rm AW}\!,\ssigma_2}_N(f)}=\cO(N^{-1})\,.
$$
\end{prop}
%+++++++++++++++++++++++++++++
%As a result, the convergence of $(H_{\psi_N}^{\ssigma})$ to the semiclassical 
%measure $\mu$ is independent of the choice of $\ssigma$.

%=====================================================
\subsubsection{Standard quantization of the baker's map}\label{s:quantum-baker2}
%=====================================================
Strictly speaking, the quantization 
of the closed baker's map $\hB_{\vr}$ is well-defined only 
if the coefficients $\vr$ are such that
\bequ\label{e:condition}
N\,r_i=N_i\in\IN,\quad i=0,1,2\,.
\eequ
Yet, in the
semiclassical limit $N\to\infty$ one can, if necessary,
slightly modify the $r_i$ by amounts $\leq 1/N$ in order to satisfy this condition: such
a modification is irrelevant for the classical dynamics. Assuming 
\eqref{e:condition}, the quantization of $\hB_{\vr}$ on $\hn$ is given by the
following unitary matrix in the position basis \cite{BV89}:
\bequ\label{e:A_N}
\hB_{\vr,N}=F_N^{-1}\begin{pmatrix}F_{N_0}&&\\&F_{N_1}&\\&&F_{N_2}\end{pmatrix}\,.
\eequ
Here $F_{N_i}$ denotes the $N_i$-dimensional discrete Fourier transform, 
\bequ\label{e:DFT}
(F_{N_i})_{jk}={N_i}^{-1/2}\,\e^{-2\i\pi jk/N_i},\quad j,k=0,\ldots N_i-1.
\eequ
Since we already have a quantization for the $\hB_{\vr}$ and the hole is the 
rectangle $H=R_1=\set{r_0\leq q<1-r_2}$, a natural choice to quantize $B_{\vr}$ is to
project on the positions $q\in [0,1)\setminus [r_0,1-r_2)$ and then apply $\hB_{\vr,N}$.
One gets the following open propagator in the position basis \cite{SaVa96}:
\bequ\label{e:B_N}
B_{\vr,N}=F_N^{-1}\begin{pmatrix}F_{N_0}&&\\&0&\\&&F_{N_2}\end{pmatrix}\,.
\eequ
This is the ``standard'' quantization of the open baker's map $B_{\vr}$. These
matrices are obviously contracting on $\hn$.
The semiclassical
connection between the matrices $\hB_{\vr,N}$ (resp. $B_{\vr,N}$) and the classical map 
$\hB_{\vr}$ (resp. $B_{\vr}$) has
been analyzed in detail in \cite{NonZw06}; this analysis implies the property
\eqref{e:axiom2-map} with respect to the Weyl quantization.
%%%%%%%%%%%%%%%%%%%%
\begin{figure}[htbp]
\begin{center}
\includegraphics[width=.6\textwidth]{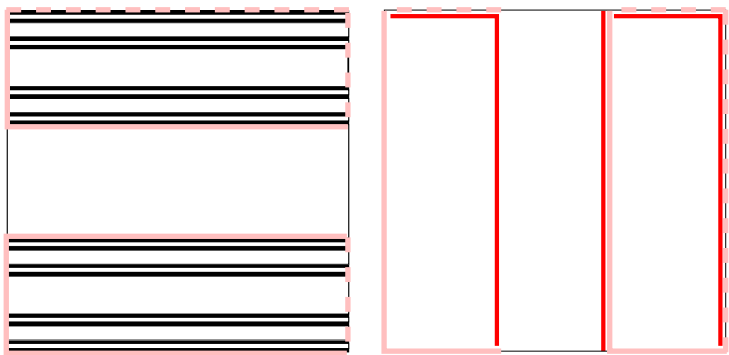}
\caption{\label{f:gamma+} On the left, we show the backward trapped set $\Gamma_+$ for the
symmetric open baker $B_{\vr_{sym}}$ (black), and the discontinuity set $D(B_{\vr_{sym}}^{-1})$ 
(pink/light gray). Their union gives $\Gamma_+\sqcup D_{-\infty}$, which contains
the support of semiclassical measures (see Thm.~\ref{thm:main}, $i)$).
The dotted lines indicate identical points on $\t2$. 
On the right, we show the backwards image 
$B_{\vr_{sym}}^{-1}(D(B_{\vr_{sym}}^{-1}))$ involved in 
Thm~\ref{thm:main}, $iii)$ (pink/light gray), and the projection on $\t2$ of the set 
$\Sigma\setminus\Sigma''$ defined in \eqref{e:Sigma''} (red/dark gray).} 
\end{center}
\end{figure}
%%%%%%%%%%%%%%%%%%%%
Below we give an alternative proof of the property \eqref{e:axiom2-map} for the open
propagator $B_{\vr,N}$, with respect to the
anti-Wick quantization. The proof uses the semiclassical propagation of 
coherent states analyzed in \cite{DENW06}. 

To start with, we precisely give the continuity set
of $B_{\vr}^{-1}$:
$$
C(B_{\vr}^{-1})=\tR_0\sqcup \tR_2\,,
$$
where $\tR_0=\set{q\in(0,1),\ p\in (0,r_0)}\subset B_{\vr}(R_0)$ 
and similarly for $\tR_2$: in the case of the symmetric baker,
these are the open grey rectangles on 
the right plot of Fig.~\ref{f:3-baker}.
On the other hand, the hole $H^{-1}=B_{\vr}(R_1)=\set{q\in [0,1),\ p\in [r_0,1-r_2)}$ (this is 
the white strip, including the vertical side). 
The discontinuity set $D(B_{\vr_{sym}}^{-1})$
is shown in Fig.~\ref{f:gamma+} (pink/light gray lines in the left plot).

Any observable 
$a\in C^\infty_c\big(C(B_{\vr}^{-1})\sqcup \ring{H}^{-1}\big)$ is supported 
at a distance $\geq\delta>0$
from the discontinuity set $D(B_{\vr}^{-1})$. 
Let select some smooth $\ssigma\in C^\infty(\t2,(0,\infty))$, and consider
the corresponding anti-Wick quantization (see \eqref{e:anti-Wick}).
From the definition \eqref{e:anti-Wick}, the operator 
${\rm Op}^{{\rm AW}\!,\ssigma}(a)$ only involves coherent states
$\psi_{\bx,\ssigma(\bx)}$ located at distance $\geq\delta>0$ from $D(B_{\vr}^{-1})$.
Adapting the proof of \cite[Prop.5]{DENW06} to the general open baker $B_{\vr}$,
one can show the following propagation for such states:
\bequ\label{e:CS-evol}
B_{\vr,N}^\dagger\; \psi_{\bx,\ssigma(\bx)}=\bbbone_{M^{-1}}(\bx)\,\e^{i \theta(\bx,N)} \,
\psi_{\bx',\ssigma'(\bx')}
+\cO_{\hn}(\e^{-N\,C(\ssigma,\del)})\,,\qquad N\to\infty\,.
\eequ
Here $\bx'=B_{\vr}^{-1}(\bx)$, $\ssigma'(\bx')=\ssigma(\bx)/r_\eps^2$ if $\bx'$ lies inside the
rectangle $R_{\eps}$, and $\theta(\bx,N)$ is a phase which can be explicitly computed. 

Through the symplectic 
change of variable $\by=B_{\vr}^{-1}(\bx)$ for $\by\in M$, one gets
$$
B_{\vr,N}^\dagger\,{\rm Op}^{{\rm AW}\!,\ssigma}(a)\,B_{\vr,N}= 
{\rm Op}^{{\rm AW}\!,\ssigma'}(\bbbone_{M}\times (a\circ B_{\vr})) +
\cO_{\cB(\hn)}(\e^{-N\,C'(\ssigma,\del)})\,.
$$
The function $\ssigma'$ is obtained by taking $\ssigma'(\by)=\ssigma(B_{\vr}(\by))/r_\eps^2$ for 
$\by\in B_{\vr}^{-1}(\supp a\cap \tR_{\eps})$, and smoothly extending the function to 
$\ssigma'\in C^\infty(\t2,(0,\infty))$.
Since the quantizations with parameters $\ssigma$ and $\ssigma'$ are equivalent 
(Prop.~\ref{p:quant}), we have proven
the property \eqref{e:axiom2-map} for the family $(B_{\vr,N})$. $\hfill\square$

%%%%%%%%%%%%%%%%%%%%%%%%%%%%%%%%%%%%%%%%%%%%%%%%%%%%%%
\section{Fractal Weyl law for the quantized open baker} 
%%%%%%%%%%%%%%%%%%%%%%%%%%%%%%%%%%%%%%%%%%%%%%%%%%%%%%
In this section, which mainly presents numerical results, 
we exclusively consider the open baker's maps $B_{\vr}$ and their 
quantizations \eqref{e:B_N}. Our aim is to investigate the precise notion of dimension
entering the fractal Weyl law conjectured below, through a numerical study which
complements the one performed in \cite{NonZw05}. Still, we believe that most
statements should hold as well if we replace $B_{\vr}$ by a map $T$ obtained by
restricting an Anosov diffeomorphism $\hT$ outside some 
``small hole'', as described in \cite{CherMar97}.

From the explicit formula \eqref{e:B_N}, the subspace of $\hn$ spanned by 
$\set{\bq_j,\ N_0\leq j<N-N_2}$ is in the kernel of $B_{\vr,N}$. We call
``nontrivial'' the spectrum of $B_{\vr,N}$ on the complementary subspace.
That spectrum is situated inside the unit disk. In the semiclassical 
limit $N\to\infty$, most of it accumulates 
near the origin, which corresponds to ``short-living'' eigenvalues \cite{SchoTwo04}.
We rather focus on ``long-living'' eigenvalues, situated in some annulus away from the origin. 
By analogy with the case of potential scattering \eqref{e:Weyl-law}, a fractal Weyl law
for the semiclassical density of ``long-living'' eigenvalues was conjectured in \cite{NonZw06}:
%%%%%%%%%%%%%%%%%%%%%%%%%%%
\begin{conj}[Fractal Weyl Law]\label{conj:FWL}
Let $B_{\vr,N}$ be the quantized open baker's map described in \S\ref{s:quantum-baker}.  
Then, for any radius $0<r<1$, there exists $C_r\geq 0$ such that
\bequ\label{e:FWL}
n(N,r)\defeq
\#\set{ \lambda \in \Spec ( B_{\vr,N} ),\ :\  |\lambda|\geq r}=C_r\,N^{d}\,+o(N^{d }),
\qquad N\to\infty\,.
\eequ
The eigenvalues
are counted with multiplicities, and $2d$ is an appropriate 
fractal dimension of the trapped set $K$. 
\end{conj}
%%%%%%%%%%%%%%%%%%%%%%%%%%%
%So far, this conjecture could be proven only for the Walsh quantization $\tB_N$
%of the symmetric baker $B_{\vr_{sym}}$, see \S\ref{s:Walsh-baker}.
%This Weyl law was numerically checked for the open kicked rotator \cite{SchoTwo04}, and
%for the ``standard'' quantum open baker $B_{\vr_{sym},N}$, especially 
%for $N$ along geometric sequences $N=N_o\,3^k$, $k\geq 1$ \cite{NonZw05}. 

%%%%%%%%%%%%%%%%%%%%%%%%%%%%%%%%%%%%%%%%%%%%%%%%%%%%%%
\subsection{Which dimension plays a role?}\label{s:infodim?} 
%%%%%%%%%%%%%%%%%%%%%%%%%%%%%%%%%%%%%%%%%%%%%%%%%%%%%%
%%%%%%%%%%%%%%%%%%%%
\begin{figure}[htbp]
\begin{center}
\includegraphics[width=8cm]{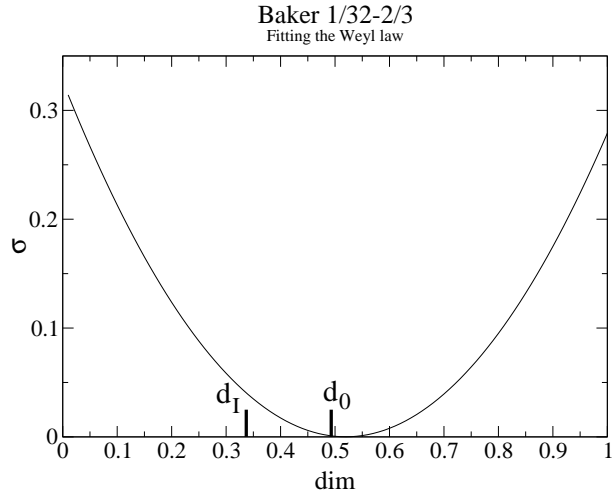}
\caption{\label{f:fit} Standard deviations when fitting the Weyl law \eqref{e:FWL} 
to various dimensions, integrated on $0.1\leq r\leq 1$. The two marks on the horizontal
axis indicate the theoretical values $d_I$ and $d_0$.}
\end{center}
\end{figure}
%%%%%%%%%%%%%%%%%%%%
%%%%%%%%%%%%%%%%%%%%
\begin{figure}[htbp]
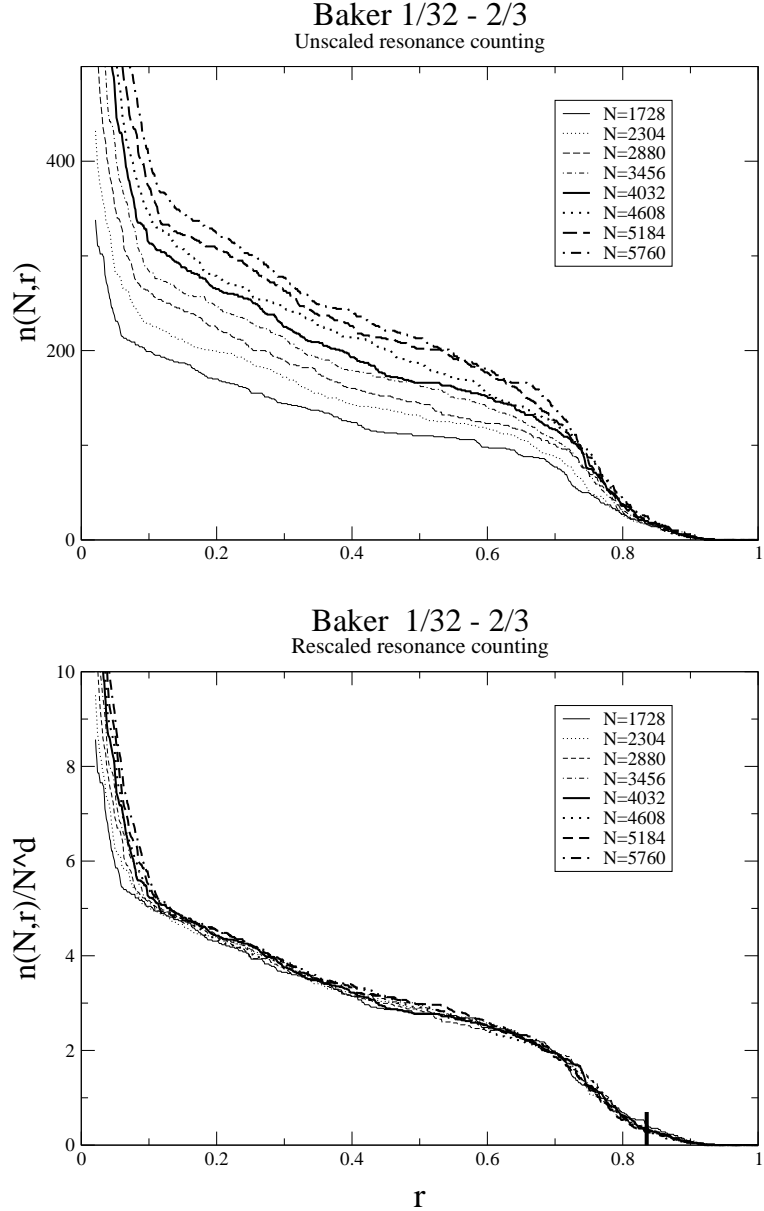

\begin{center}
\includegraphics[width=10cm]{spect-asymB-unscaled-BW.eps}
\includegraphics[width=10cm]{spect-asymB-rescaled-BW2.eps}
\caption{\label{f:asymB}Top: spectral counting function for the asymmetric baker $B_{\vr_{asym}}$, 
for various values of Planck's constant $N$. Bottom: same curves vertically
rescaled by $N^{-d_0}$. The thick tick mark indicates the radius $\sqrt{\Lambda_{nat}}$ 
corresponding to the natural measure.}
\end{center}
\end{figure}
%%%%%%%%%%%%%%%%%%%%
In the proofs for upper bounds of the Weyl law \eqref{e:Weyl-law}, 
the exponent $d$ is defined in terms of the 
{\em  upper Minkowski dimension} of the trapped set $K$ \cite{Sjo90,Zw99,GLZ04,SjoZw05}. 
In the case of the open baker $B_{\vr}$, we therefore expect that
the exponent $d$ appearing in the conjecture (resp. $d+1$) is given by the
Minkowski dimension of the Cantor set $\cC_{\vr}$ (resp. $\Gamma_+$), which is equal to
its box, Hausdorff and packing dimensions. We call this theoretical
value $d_0$.

For the symmetric baker $B_{\vr_{sym}}$, the Hausdorff dimension $d_H(\Gamma_+)=d_0+1$ 
happens to be equal to the Hausdorff dimension of the
natural measure $\mu_{nat}$, defined by
$$
d_H(\mu_{nat})=\inf_{A\subset\t2,\ \mu_{nat}(A)=1} d_H(A)\,.
$$
For a nonsymmetric baker's map $B_{\vr}$, those two Hausdorff dimensions 
only satisfy the inequality
$d_H(\mu_{nat})\leq d_H(\Gamma_+)$ (see the explicit expressions below).
We want to investigate a possible ``role'' of the natural eigenmeasure $\mu_{nat}$
regarding the structure of the quantum spectrum. It is therefore legitimate to
ask the following 
\begin{question}\label{qu0}
Is the correct exponent in the Weyl law \eqref{e:FWL}
given by $d_I=d_H(\mu_{nat})-1$ instead of $d_0=d_H(\Gamma_+)-1$?
\end{question}
Here the suffix $I$ indicates that $d_I$ is sometimes called
the {\em information dimension} \cite{HJPPS86}.
As mentioned above, both dimensions are equal for the symmetric baker $B_{\vr_{sym}}$,
for which the Weyl law \eqref{conj:FWL} has been numerically tested in \cite{NonZw05}.
They are also equal in the case of a closed map on $\t2$:
in that case, the Weyl law has exponent $1$ (the whole spectrum
lies on the unit circle), and we have $d_H(\t2)=d_H(\mu_{L})=2$. 

For a nonsymmetric baker $B_{\vr}$, the two dimensions
take different values:
$$
d_0\quad \text{is the solution of}\quad r_0^d+r_2^d=1,\qquad\text{while}\quad
d_I=\frac{r_0\log p_0+r_2\log p_2}{r_0\log r_0+r_2\log r_2}\,.
$$
To answer the above question, we
considered a very asymmetric baker, taking $\vr_{asym}$ with $r_0=1/32$, $r_2=2/3$.
The two dimensions then
take the values $d_0\approx 0.493$, $d_I\approx 0.337$. We computed the counting function
$n(N,r)$ for several radii $0.1\leq r\leq 1$ and several values of $N$. We then tried
to fit the Weyl law \eqref{e:FWL} with an exponent $d$ varying in a certain range,
and computed the standard deviations (see Fig.~\ref{f:fit}). 
The numerical result is unambiguous: the best fit clearly occurs away from $d_I$, 
but it is close to $d_0$. This numerical test rules out the possibility 
that $d_I$ provides the correct exponent of the Weyl law, and suggests to 
indeed take $d=d_0$.

To further illustrate the Weyl law for the asymmetric baker $B_{\vr_{asym}}$,  
we plot in Fig.~\ref{f:asymB} (top) the counting functions
$n(N,r)$ as a function of $r\in(0,1)$, for several values of $N$.
On the bottom plot, we rescale $n(N,r)$ by the power $N^{-{d_0}}$: 
the rescaled curves almost perfectly
overlap, indicating that the scaling \eqref{e:FWL} is correct.
\begin{rem}\label{r:fill}
On figure~\ref{f:asymB} (right) the rescaled 
counting function seems to converge to a function which is strictly decreasing
on an interval $[\lambda_{\min},\lambda_{\max}]$, where $\lambda_{\min}\approx 0.1$, 
$\lambda_{\max}\approx 0.9$. This implies that the spectrum of $B_{\vr_{asym},N}$ becomes
dense in the whole annulus $\set{\lambda_{\min}\leq|\lambda|\leq\lambda_{\max}}$, 
when $N\to\infty$. 
Therefore, at this heuristic level, 
for any $\lambda\in [\lambda_{\min},\lambda_{\max}]$
one may consider sequences of 
eigenvalues $(\lambda_N)_{N\geq 1}$ with the property $|\lambda_N|\Nto8 \lambda$.
In particular, we may consider sequences converging to $\sqrt{\Lambda_{nat}}$.
\end{rem}

%%%%%%%%%%%%%%%%%%%%%%%%%%%%%%%%%%%%%%%%%%%%%%%%%%%%%%%%%%%%%%%%%%%%%%%
\section{Localization of resonant eigenstates}\label{s:localization}
%%%%%%%%%%%%%%%%%%%%%%%%%%%%%%%%%%%%%%%%%%%%%%%%%%%%%%%%%%%%%%%%%%%%%%%

We now return to the general framework of an open map $T$
on a subset $M$ of a compact phase space $\hM$, which is
quantized into a sequence of operators $(T_N)_{N\to\infty}$ according to Axioms~\ref{d:axiom}.
We want to study the semiclassical measures associated with the long-living eigenstates of $T_N$.

To start with, we fix some $\lambda_m\in(0,1)$, such that for $N$ large enough, 
$\Spec T_N\cap \set{|\lambda|\geq \lambda_m}\neq \emptyset$. We
can then consider sequences of eigenstates $(\psi_N)_{N\to\infty}$ such that, for $N$
large enough, 
\bequ\label{e:eigenstates}
T_N\,\psi_N=\lambda_N\,\psi_N\,,\qquad \norm{\psi_N}=1\,,\qquad 
|\lambda_N|\geq \lambda_{m}\,.
\eequ
The role of the (quite arbitrary) lower bound $\lambda_{m}>0$ is to ensure 
that the eigenstates we consider are ``long-living''.
Up to extracting a subsequence, we can assume that
$(\psi_N)$ converges to a certain semiclassical measure $\mu$ 
(see \S\ref{s:semicl-measure}). 

\medskip

To state our result, we first
we need to analyze the continuity sets of the backward iterates of $T$. 
For any $n\geq 1$, the map $T^{-n}$ is defined on the set $M^{-n}$. Its 
continuity set $C(T^{-n})$ can be obtained iteratively through 
$$
C(T^{-n})=T\big(C(T^{-n+1})\cap C(T)\big)\,,\quad n\geq 2\,.
$$
The set $M_{-n}$ of points escaping exactly at 
time $(-n)$ can also be split between its continuity subset
its continuity subset
$$
CM_{-n}\defeq C(T^{-n+1})\cap \ring{M}_{-n}=T^{n-1}(C(T^{n-1})\cap \ring{H}^{-1})\,,
$$
which is a union of open connected sets, and its discontinuity subset
$DM_{-n}=M_{-n}\setminus CM_{-n}$, which has Minkowski content zero (in the case $n=1$,
we take $CM_{-1}=\ring{H}^{-1}$).
Using this splitting of the $M_{-n}$, the decomposition \eqref{e:split2} can be recast into:
\bequ\label{e:decompo-final}
\hM=C_{-\infty}\sqcup D_{-\infty} \sqcup\Gamma_+,\quad\text{where}\quad
C_{-\infty}=\big(\bigsqcup_{n=1}^\infty CM_{-n}\big),\quad
D_{-\infty}=\big(\bigsqcup_{n=1}^\infty DM_{-n}\big)\,.
\eequ
The set $C_{-\infty}$ consists of points
which eventually fall in the hole when evolved through $T^{-1}$, 
and remain at finite distance from $D(T^{-1})$ all along their transient trajectory.

We can now state our main result concerning semiclassical measures.
%%%%%%%%%%%%%
\begin{thm}\label{thm:main}
Assume that a sequence of eigenstates $(\psi_N)_{N\to\infty}$ of the 
open quantum map $T_N$, with eigenvalues $|\lambda_N|\geq\lambda_m>0$,
converges to the semiclassical measure $\mu$ on $\hM$. Then the following hold.

i) the support of $\mu$ is a subset of $\Gamma_+\sqcup D_{-\infty}$.

ii) If $\mu\big(C(T^{-1})\big)>0$, there exists $ \Lambda\in [\lambda_{m}^2,1]$ such
that the eigenvalues $(\lambda_N)_{N\to\infty}$ satisfy
$$
|\lambda_N|^2\xrightarrow{N\to\infty} \Lambda\,.
$$ 
For any Borel subset $S$ not intersecting 
$D(T^{-1})$, one has
$\mu(T^{-1}(S))=\Lambda\,\mu(S)$. 

iii) If $\mu(D(T^{-1}))=\mu\big(T^{-1}(D(T^{-1}))\big)=0$, 
then $\mu$ is an eigenmeasure of $T$, with decay rate $\Lambda$. 
\end{thm}
%%%%%%%%%%%%%
\begin{proof}
To prove the first statement, let us choose some $n\geq 1$, and take an observable
$a\in C^\infty_c(CM_{-n})$. Every point $x\in\supp a$ has the property that
for any $0\leq j <n-1$, $T^{-j}(x)\in C(T^{-1})$, while $T^{-n+1}(x)\in \ring{H}^{-1}$.
Applying iteratively the property \eqref{e:axiom2-map}, one finds that in the 
semiclassical limit,
$$
(T_N^{\dagger})^{n}\,\Op_N(a)\,(T_N)^{n}\sim 0\,.
$$
We take into account the fact that $\psi_N$ is a right eigenstate of $T_N$:
$$
\la \psi_N,\,\Op_N(a)\,\psi_N\ra=
|\lambda_N|^{-2n}\,\la \psi_N,(T_N^{\dagger})^{n}\,\Op_N(a)\,(T_N)^{n}\,\psi_N\ra\,.
$$
Using $|\lambda_N|\geq \lambda_m$, these two expressions imply
$\mu(a)\defeq \int a\,d\mu=0$.
Since $n$ and $a$ were arbitrary, this shows $\mu(C_{-\infty})=0$, which is the first statement.

From the assumption in $ii)$, we may select $a\in C^\infty_c(C(T^{-1}))$ 
such that $\int a\,d\mu>0$. The first iterate $a\circ T$ is supported in $C(T)\subset M$.
Applying \eqref{e:axiom2-map}, we get
\bequ\label{e:eigenstate}
|\lambda_N|^{2}\,\la\psi_N,\,\Op_N(a)\,\psi_N\ra \sim 
\la \psi_N,\,\Op_N(a\circ T)\,\psi_N\ra\,.
\eequ
For $N$ large enough, the matrix element
$\la\psi_N,\,\Op_N(a)\,\psi_N\ra > \mu(a)/2$, so we may divide the
above equation by this element, and obtain
$$
|\lambda_N|^2\Nto8 \frac{\mu(a\circ T)}{\mu(a)}\,.
$$
We call $\Lambda\geq |\lambda_m|^2$ this limit, which is obviously 
independent of the choice of $a$. For any Borel set
$S\subset C(T^{-1})$, one can approximate $\bbbone_{S}$ by smooth functions supported
in $C(T^{-1})$, to prove that 
$$
\mu(T^{-1}(S))=\Lambda\,\mu(S)\,.
$$
If $S\subset \ring{H}^{-1}$, we know that $\mu(S)=0$ and $T^{-1}(S)=\emptyset$ ,
so the above equality still makes sense. This proves $ii)$.

To obtain $iii)$, we
split any Borel set $S$ into $S=(S\cap D(T^{-1}))\sqcup CS$.
From $ii)$, we have $\mu(T^{-1}(CS))=\Lambda\,\mu(CS)$.
By assumption, $\mu(S\cap D(T^{-1}))=\mu(T^{-1}(S\cap D(T^{-1})))=0$, so
we get $\mu(T^{-1}(S))=\Lambda\,\mu(S)$.
\end{proof}

%%%%%%%%%%%%%%%%%%%%%%%%%%%%%%%%%%%%%%%%%%%%%%%%%%%%%%%%%%%%%%%%%%%%%%%
\subsection{Semiclassical measures of the open baker's map} \label{s:semiclass}
%%%%%%%%%%%%%%%%%%%%%%%%%%%%%%%%%%%%%%%%%%%%%%%%%%%%%%%%%%%%%%%%%%%%%%%

In this section we apply the above theorem to the case of some open baker's map $B_{\vr}$,
quantized as in \S\ref{s:quantum-baker2}. We also numerically compute the Husimi
measures $H^1_{\psi_N}$ associated with some quantum eigenstates (we choose the isotropic
squeezing $\ssigma=1$ by convenience): although we cannot really go to the semiclassical 
limit, we hope that for $N\sim 1000$ these Husimi measures already give some idea of the
semiclassical measures.

\subsubsection{Applying Theorem~\ref{thm:main}}

To simplify the presentation we will restrict the discussion to the symmetric baker 
$B_{\vr_{sym}}$, which 
will be denoted by $B$ in short. The discontinuity set $D(B^{-1})$, its backwards image 
and the trapped sets were described in \S\ref{s:baker-trapped} and \S\ref{s:quantum-baker2}, and
plotted in Fig.~\ref{f:gamma+}.
The sets $M_{-n}$ and their continuity subsets are also simple to describe using symbolic dynamics.
For any $n=1$, we know that $CM_{-1}=\ring{H}^{-1}$ is the open rectangle 
$\set{q\in [0,1),\ p\in (1/3,2/3)}$. 
For $n\geq 2$, 
$CM_{-n}$ is the union of $2^{n-1}$ horizontal rectangles, indexed by the $n$-sequences 
$\bep=\ep_{-1}\ldots\ep_{-n}$ such that $\ep_{-i}\in\set{0,2}$, $i=1,\ldots,n-1$, while
$\ep_{-n}=1$. Each such rectangle is
of the form $\set{q\in (0,1),\ p\in (p(\bep), p(\bep)+3^{-n})}$, where 
$p(\bep)\equiv \cdot\,\ep_{-1}\ldots\ep_{n}$ in ternary decomposition. 
Some of those  rectangles are shown in Fig.~\ref{f:Cantor} (center).
The union of all these rectangles (for $n\geq 1$) makes up
$C_{-\infty}$. Its complement $\Gamma_+\sqcup D_{-\infty}$ is given by
$$
\Gamma_+\sqcup D_{-\infty}=\big([0,1)\times \cC_{\vr_{sym}}\big)\cup D(B^{-1})\,.
$$
This set is shown in Fig.~\ref{f:gamma+} (left).
%%%%%%%%%%%%%%%%%%%%
\begin{figure}[htbp]
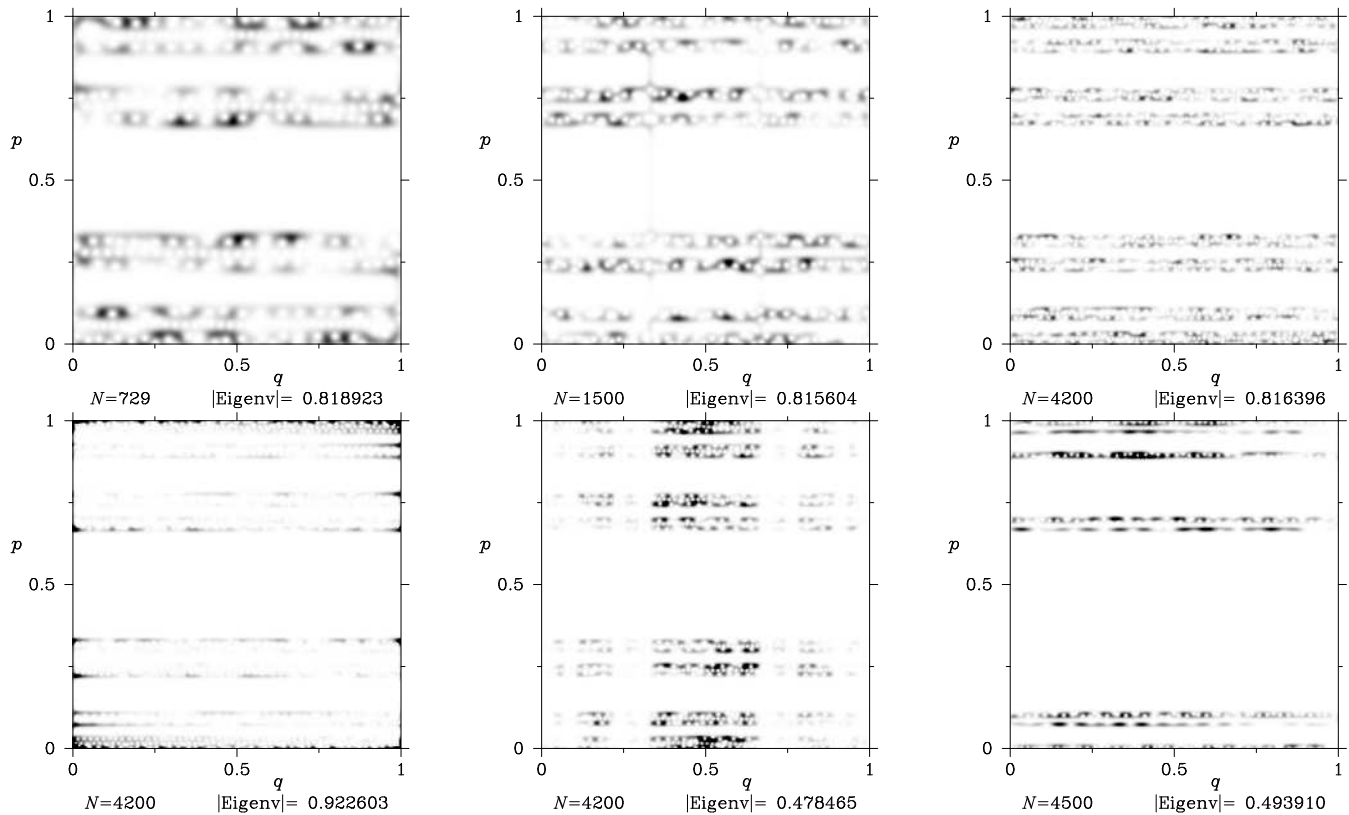

\begin{center}
\rotatebox{-90}{\includegraphics[width=.33\textwidth]{3states-nat.ps}}
%\rotatebox{-90}{\includegraphics[width=6cm]{3states-sym.ps}}
\rotatebox{-90}{\includegraphics[width=.33\textwidth]{3states-2maps.ps}}
%\rotatebox{-90}{\includegraphics[width=6cm]{2states9-3.ps}}
\caption{\label{f:eigenstates} Husimi densities of (right) 
eigenstates of $B_{\vr,N}$ (black=large values, white=0).
Top: 3 eigenstates of $B_{\vr_{sym},N}$ with $|\lambda_N|\approx \sqrt{\Lambda_{nat}}$.
Bottom left, center: two eigenstates of $B_{\vr_{sym},N}$ with different $|\lambda_N|$. 
Bottom right: one eigenstate of $B_{\vr_2,N}$.}
\end{center}
\end{figure}
%%%%%%%%%%%%%%%%%%%%

In Figure~\ref{f:eigenstates}, we plot the Husimi densities $H^1_{\psi_N}$ of some
(right) eigenstates of $B_{\vr,N}$, for the symmetric baker and 
an asymmetric one, $\vr_{2}=(1/9,5/9,1/3)$.

\begin{rem}
We notice that all Husimi functions are indeed
very small in the horizontal rectangles $M_{-n}$ for $n=1,2$ (in case $N\leq 1500$),
and also $n=3$ for $N=4200$. Using \eqref{e:CS-evol}, 
one can refine the proof of Thm.~\ref{thm:main} to show that $H^1_{\psi_N}(x)=\cO(\e^{-cN})$ 
for $x\in C_{-\infty}$, $N\to\infty$. 
\end{rem}

\begin{rem}\label{r:rem2}
Although this is not proven in our theorem, all the Husimi functions we have computed
are very small on the set $D(B^{-1})\setminus \Gamma_+\subset \set{q=0}$
(see the left plot in Fig.~\ref{f:gamma+}). 
On the other hand, some of these Husimi functions are large on
$D(B^{-1})\cap \Gamma_+$,
so we cannot rule out the possibility that 
some semiclassical measures $\mu$ charge this set. 
For instance, in some of the Husimi plots ($e.g.$
the bottom left in Fig.~\ref{f:eigenstates}), we clearly see
a strong peak at the origin, and
lower peaks on other points of $D(B^{-1})\cap \Gamma_+$.
We call ``diffractive'' the components of an eigenstate localized on $D_{-\infty}$. 
\end{rem}

From these observations, we state the following
\begin{conj} \label{c:eigenm}
Let $B_{\vr}$ be an open baker's map, quantized by \S\ref{s:quantum-baker2}. Then,
\begin{itemize} 
\item all long-living semiclassical measures are supported on 
$\overline{\Gamma_+}$
\item ``almost all'' long-living eigenstates are non-diffractive, 
that is, their weight
on $D(B^{-1})$ and $B^{-1}(D(B^{-1}))$ are negligible
in the semiclassical limit. 
The corresponding semiclassical measures are then eigenmeasures of $B_{\vr}$.
\end{itemize}
\end{conj}
In the next section we further comment on the above Husimi plots.
%%%%%%%%%%%%%%%%%%%%%%%%%%%%%%%%%%%%%%%%%%%%%%%%%%
\subsection{Abundance of semiclassical measures}\label{s:abundance} 
%%%%%%%%%%%%%%%%%%%%%%%%%%%%%%%%%%%%%%%%%%%%%%%%%%

Let us draw some consequences from Theorem~\ref{thm:main} and the following remarks.
Statement $ii)$ of the theorem strongly constrains the 
converging sequences of eigenstates: a sequence $(\psi_N)$ can
converge to some measure $\mu$ (with $\mu(C(B^{-1})) > 0$) only if the
corresponding eigenvalues $(\lambda_N)$ asymptotically approach the circle of radius 
$\sqrt{\Lambda}$. 

We take for granted the density argument in Remark~\ref{r:fill} and assume that
a ``dense interval'' $[\lambda_{\min},\lambda_{\max}]\subset (0,1]$
exists for any open baker's map $B_{\vr}$. Hence,
for any $\Lambda\in [\lambda_{\min}^2,\lambda_{\max}^2]$ there exist
many sequences of eigenstates of $B_{\vr,N}$, 
such that $|\lambda_N|^2\Nto8 \Lambda$. From our Conjecture~\ref{c:eigenm}, almost any
semiclassical measure associated with such a sequence will be an eigenmeasure with
decay rate $\Lambda$. Therefore, two converging sequences $(\psi_N)$, $(\psi_N')$
associated with limiting decays $\Lambda\neq\Lambda'$ will necessarily converge
to different eigenmeasures. This already shows that the semiclassical eigenmeasures
generated by all possible sequences in the annulus 
$\set{\lambda_{\min}\leq|\lambda|\leq\lambda_{\max}}$ form an uncountable family.

According to \S\ref{s:eigenmeasures}, for each decay rate $\Lambda\in (0,1)$,
there exist uncountably many eigenmeasures. A natural question thus concerns the
variety of semiclassical measures associated with a given 
$\Lambda\in [\lambda_{\min}^2,\lambda_{\max}^2]$:

\begin{question}\label{qu1}$\ $

For a given $\Lambda\in [\lambda_{\min}^2,\lambda_{\max}^2]$, what are the
semiclassical measures of $B_{\vr}$ of decay rate $\Lambda$?
\begin{itemize}
\item is there a unique such measure? 
\item otherwise, is some limit measure  ``favored'', in the sense that ``almost all'' sequences
$(\psi_N)$ with $|\lambda_N|^2\to \Lambda$ converge to $\mu$?
\item can the {\em natural measure} $\mu_{nat}$ be obtained as a semiclassical measure?
\end{itemize}
\end{question}
The same type of questions were asked by Keating and coworkers in ref.~\cite{KNPS06}. 
At present we are unable to answer them rigorously for the quantum open baker $B_{\vr,N}$. 

From a heuristic point of view we notice the following features 
in the plots of Fig.~\ref{f:eigenstates}. The three
top plots correspond to eigenvalues $|\lambda_N|$ close to the value 
$\sqrt{\Lambda_{nat}}\approx 0.8165$. However, the Husimi measures on the left and center
seem very different from the natural measure (the latter is approximated by the black rectangles
in Fig.~\ref{f:gamma+}). These two Husimi functions seem to charge different 
parts of $\Gamma_+$. Only the rightmost state seems compatible with a convergence 
to $\mu_{nat}$.

As commented in Remark~\ref{r:rem2}, the bottom-left state has strong concentrations on
$D(B^{-1})\cap\Gamma_+$. For the various values of $N$ we have investigated, this 
concentration seems characteristic of the eigenstate with the largest 
$|\lambda_N|$. The discontinuities of $B_{\vr}$ manage to ``trap'' those quantum states 
better than periodic orbits on $C(B^{-1})\cap\Gamma_+$. Such states seem ``very diffractive''.

Finally, the bottom-center state, with a smaller eigenvalue, clearly
shows a selfsimilar structure along the horizontal direction, with a
probability inside the hole higher than for the top eigenstates. 
This feature had already been noticed in \cite{KNPS06} for averages over Husimi 
functions with comparable $|\lambda_N|$.

In section~\ref{s:walsh} we will address Questions~\ref{qu1} for
a different quantization of the symmetric open baker, namely 
the Walsh-quantized open baker, where one can compute some semiclassical
measures explicitly.

Before that, we explain how to construct approximate eigenstates
({\em pseudomodes}) for the quantum baker $B_{\vr_{sym},N}$.

%%%%%%%%%%%%%%%%%%%%%%%%%%
\subsection{Pseusomodes and pseudospectrum} 
%%%%%%%%%%%%%%%%%%%%%%%%%%

From the explicit representation of eigenmeasures given in Proposition~\ref{p:eigen-decompo},
it is possible to construct approximate eigenstates of $T_N$ by 
backward propagating wavepackets localized on the set $\Gamma_+^{(1)}=\Gamma_+\cap H$.

We call these approximate eigenstates 
{\it pseudomodes}, by analogy with the recent literature on nonselfadjoint semiclassical
operators (see e.g. \cite{DSZ04,BU03}). Those papers deal with pseudodifferential operators
obtained by quantizing complex-valued observables, and they construct
pseudomodes of error $\cO(\hbar^\infty)$, microlocalized at a single phase space point.

Our pseudomodes will be less precise and will be microlocalized on a countable set. 
We will restrict ourselves to the case of the
open symmetric baker $B=B_{\vr_{sym}}$, but our construction works for 
any $B_{\vr}$, and can probably be extended to other maps or phase spaces.

Inspired by the pointwise eigenmeasures \eqref{e:mu-traject}, we construct 
approximate eigenstates by backwards evolving coherent states localized in $\Gamma_+^{(1)}$.
Precisely, for any $\bx_0\in \Gamma_+^{(1)}$, $\ssigma>0$ and $\lambda\in \IC$, $|\lambda|<1$, 
$N\in\IN$, we define the following quantum state:
\begin{equation}\label{e:quasim}
\Psi^\ssigma_{\lambda,x_0}\defeq \sqrt{1-|\lambda|^2}\;
\sum_{n\geq 0}\lambda^n\,B_{N}^{\dagger n}\,\psi_{\bx_0,\ssigma}\,,
\end{equation}
where $\psi_{\bx_0,\ssigma}$ is the coherent state defined in \S\ref{s:Husimi}, and $B_N$ is
the standard quantization of $B$.
%++++++++++++++++
\begin{prop}
Consider the symmetric open baker $B=B_{\vr_{sym}}$ and its quantization $(B_N)$.
Fix $\lambda\in\IC$ with $|\lambda|<1$. 

$i)$ Choose $\vareps>0$ small and call $\alpha=(1-\vareps)\frac{\log 1/|\lambda|}{\log 3}$. 
For any $N\in\IN^*$, 
one can choose a point $x_0(N)\in \Gamma_+^{(1)}$ and a squeezing 
parameter $\ssigma=\ssigma(N)>0$ such that the 
states\\ $\big(\Psi_N\defi \Psi^{\ssigma(N)}_{\lambda,x_0(N)}\big)_{N\to\infty}$ defined
in \eqref{e:quasim} satisfy
$$
\norm{\Psi_N}=1+\cO(N^{-\alpha}),\qquad 
\norm{(B_N-\lambda)\Psi_N}=\cO(N^{-\alpha})\,,\qquad N\to\infty.
$$
$ii)$ for any $x_0\in \Gamma_+^{(1)}$, one can 
select the points $\big(x_0(N)\big)$ such that
$x_0(N)\Nto8 x_0$. In that case, the sequence $(\Psi_N)_{N\to\infty}$
converges to the eigenmeasure $\mu_{x_0,\Lambda}$ described in \eqref{e:mu-traject},
with $\Lambda=|\lambda|^2$.
\end{prop}
%++++++++++++++++
This proposition shows that, for
any $\alpha>0$ and $N$ large enough,
the $N^{-\alpha}$-pseudospectrum of $B_N$ contains the disk 
$\set{|\lambda|\leq 3^{-\alpha(1+\vareps)}}$.
We notice that the errors are not very small, and increase when $|\lambda|\to 1$. We should
emphasize that, in our numerical trials, we never found eigenstates of $B_N$ with
Husimi measures looking like $\mu_{x_0,\Lambda}$.

\subsubsection*{Proof of the proposition}
To control the series \eqref{e:quasim}, 
we would like to ensure that the evolved
coherent state $B_{N}^{\dagger j}\,\psi_{x_0(N),\ssigma(N)}$
remains close to an approximate coherent state 
$\psi_{x_{-j},\ssigma_{-j}}$ (as in \eqref{e:CS-evol})
up to large times $j$. 
For this we need the points $x_{-j}=B^{-j}(x_0(N))$ to stay ``far'' 
from the discontinuity set
$D(B^{-1})$, and we also need all $\psi_{x_{-j},\ssigma_{-j}}$ to be microlocalized
in a small neighbourhood of $x_{-j}$. 
Due to the hyperbolicity of $B^{-1}$, the second condition constrains
the times $j$ to be smaller than the {\it Ehrenfest time} \cite{DENW06}
\bequ\label{e:Ehrenf}
n_E=(1-\vareps)\frac{\log N}{\log 3}\,.
\eequ
Here $\vareps>0$ is the small parameter in the statement of the proposition.

To identify a good ``starting point'' $x_0(N)$, we set $n=[n_E]$, 
and consider the following subset of $\t2$, for $\delta,\gamma>0$:
$$
\cD_{n,\delta,\gamma}\defi\set{(q,p)\in\t2,\ q\in (1/3+\delta,2/3-\delta),\ 
\forall k\in\IZ,\ |p-\frac{k}{3^n}|>\gamma}\cap\Gamma^{(1)}_+\,.
$$
We first show that this set is nonempty if $\delta<1/9$ and 
$\gamma=3^{-n}\gamma'$, $\gamma'<1/9$.
Take $q\in (1/3+\delta,2/3-\delta)$ arbitary, and select $p=p(\bep)$ with 
the following properties: take all indices $\ep_{-j}\in\{0,2\}$, $j\geq 1$, 
so that $p\in \cC_{\vr_{sym}}$,
and require furthermore that the word $\ep_{-n-1}\ep_{-n-2}\in\set{02,20}$. 
The point $(q,p)$ is then in $\cD_{n,\delta,\gamma}$. 

Let us take any point $x_0(N)$ in that set. It automatically lies in 
$C(B^{-n})$, so its backwards iterates stay away from 
$D(B^{-1})$ at least until the time $n$.
Let us select the squeezing parameter
$\ssigma(N)=s_0=N^{-1+\vareps}$. A simple adaptation
of \cite[Prop.5]{DENW06} implies the following estimate:
\bequ\label{e:semiclass1}
\exists c,C>0,\ \ \forall j,\  0\leq j\leq n_E,\qquad
\norm{B_{N}^{\dagger j}\,\psi_{\bx_0,\ssigma_0}-\e^{\i\theta_j}\,\psi_{x_{-j},\ssigma_{-j}}}\leq
C\,\e^{-c\,N^\vareps}\,.
\eequ
Here we can take $c=\min(\delta^2,\gamma^{\prime 2})$, and $C$
is uniform w.r.to the initial point $x_0$.
From the values of $x_0$, $\ssigma_0$, one checks that 
the components 
$\la \bq_k,\psi_{x_0,\ssigma_0}\ra$ for $k/N\not\in [1/3,2/3]$ 
are of order $\cO(\e^{-cN^\vareps})$: 
that state is (very) localized in the hole $H=R_1$, so
$$
B_{N}\,\psi_{\bx_{0},\ssigma_{0}}= \cO(\e^{-c\,N^\vareps})\,.
$$
Similarly,
for any $j\leq n_E$, the state $\psi_{x_{-j},\ssigma_{-j}}$ is localized inside a certain
connected component of $M_{j+1}$ (one of the the pink/grey rectangles in Fig.~\ref{f:Cantor}, left). 
In particular, the components of 
$\la \bq_k,\psi_{x_{-j+1},\ssigma_{-j+1}}\ra$ are exponentially small in $H$. Therefore,
$$
\forall j,\  0\leq j\leq n_E,\qquad 
B_{N}\,B^\dagger_{N}\psi_{\bx_{-j},\ssigma_{-j}}= 
\psi_{\bx_{-j},\ssigma_{-j}}+\cO(\e^{-c\,N^\vareps})\,.
$$
Summing all terms $j\leq n_E$ in \eqref{e:quasim} 
and estimating the remaining series using $\norm{B_N}\leq 1$, we obtain
$$
\norm{(B_{N}-\lambda)\,\Psi^\ssigma_{\lambda,x_0}}= \cO(|\lambda|^{n_E})\,.
$$
From the definition of $n_E$, we have $|\lambda|^{n_E}=N^{-\alpha}$.

The asymptotic normalization of $\Psi^\ssigma_{\lambda,x_0}$ is proven by estimating the
overlaps between coherent states $\psi_{x_{-j},\ssigma_{-j}},\psi_{x_{-j'},\ssigma_{-j'}}$
for $j,j'\leq n_E$. Because the sets $M_{j+1}$
and $M_{j'+1}$ are disjoint for $j\neq j'$, the above mentioned localization
properties imply that
$$
\forall j,j'\leq n_E,\qquad
\la \psi_{x_{-j},\ssigma_{-j}},\psi_{x_{-j'},\ssigma_{-j'}}\ra=\delta_{j',j}+\cO(\e^{-c\,N^\vareps}),
$$
and the normalization estimate follows.
This achieves the proof of $i)$.

To prove $ii)$, let us consider an arbitrary $x_0=(q_0,p_0)\in \Gamma^{(1)}_+$. 
If $q_0\not\in\set{1/3,2/3}$,
we take $\delta$ small enough such that $q_0\in (1/3+\delta,2/3-\delta)$ and set $q_0(N)=q_0$.
Otherwise, we may let $\delta$ slowly decrease with $N$, and find 
a sequence $\big(q_0(N)\big)$ such
that $q_0(N)\in (1/3+\delta(N),2/3-\delta(N))$ and $q_0(N)\Nto8 q_0$.
On the other hand, 
$p_0\in\cC_{\vr_{sym}}=p(\bep)$ for a certain sequence $\bep$, with all 
$\ep_{-j}\in\set{0,2}$. For each $N$, we
inspect the word $\ep_{-n-1}\ep_{-n-2}$ of that sequence, where $n=[n_E]$ (see \eqref{e:Ehrenf}). 
If this word is in the set $\set{02,20}$ we keep $p_0(N)=p_0$, otherwise we 
replace this word by $02$ (keeping the other symbols unchanged) 
to define $p_0(N)$. The point $x_0(N)=(q_0(N),p_0(N))$ is then in 
$\cD_{n,\delta,\gamma}$, and 
$x_0(N)\Nto8 x_0$.

The convergence to the measure $\mu_{x_0,\Lambda}$ is due to
the localization of the coherent states $\psi_{x_{-j}, s_{-j}}$ for $j\leq n_E$.

$\hfill\square$

%%%%%%%%%%%%%%%%%%%%%%%%%%%%%%%%%%%%%%%%%%%%%%%%%%%%%%%%%%%%%%%%%%%%%%%
%%%%%%%%%%%%%%%%%%%%%%%%%%%%%%%%%%%%%%%%%%%%%%%%%%%%%%%%%%%%%%%%%%%%%%%
\section{A solvable toy model: Walsh quantization of the open baker}\label{s:walsh}
%%%%%%%%%%%%%%%%%%%%%%%%%%%%%%%%%%%%%%%%%%%%%%%%%%%%%%%%%%%%%%%%%%%%%%%
%%%%%%%%%%%%%%%%%%%%%%%%%%%%%%%%%%%%%%%%%%%%%%%%%%%%%%%%%%%%%%%%%%%%%%%
In this section we study an alternative quantization of the open symmetric
baker $B_{\vr_{sym}}$, introduced in \cite{NonZw05,NonZw06}. 
A similar quantization
of the closed baker $\hB_{\vr_{sym}}$ was proposed and 
studied in \cite{SchaCav00,TraSco02,ErmSara06},
mainly motivated by research in quantum computation.
From now on, we will drop the
index $\vr_{sym}$ from our notations, and call $B=B_{\vr_{sym}}$.

A ``simplified'' quantization of $B$ was introduced in \cite{NonZw05,NonZw06}, 
as a $N\times N$ matrix $\tB_N$, obtained by only keeping the ``skeleton''  
of $B_N$ (see \eqref{e:B_N}). Although $\tB_N$ can be defined for any 
$N$, its spectrum can be explicitly computed only 
if $N=3^k$ for some $k\in\IN$.
As explained in those references, $\tB_N$ can be interpreted as
the ``Weyl'' quantization of a  
multivalued map $\tB$ built upon $B$. In the present work we will stick to a different
interpretation of $\tB_N$, valid in the case $N=3^k$: one can then introduce
a (Walsh) quantization for observables on $\t2$, which is not equivalent with
the anti-Wick quantization of \S\ref{s:Husimi}. We will check below that
the matrices $\tB_N$ satisfies property \eqref{e:axiom2-map} with respect to that
quantization and the open baker $B$. 
Finally, this Walsh
quantization is also suited to quantize observables on
the symbol space $\Sigma$: the matrices $\tB_N$ are then quantizing the
open shift $\sigma$. We will see that this interpretation is more ``convenient'', because
it avoids problems due to discontinuities. 

We first recall the definition of the
Walsh quantization of observables on $\t2$ (or $\Sigma$), and the associated
Walsh-Husimi measures.

%%%%%%%%%%%%%%%%%%%%%%%%%%%%%%%%%%%%%%%%%%%%%%%%%%%%%%%%%%%%%%%%%%%%%%%
\subsection{Walsh transform and coherent states} \label{s:walsh0}
%%%%%%%%%%%%%%%%%%%%%%%%%%%%%%%%%%%%%%%%%%%%%%%%%%%%%%%%%%%%%%%%%%%%%%%

\subsubsection{Walsh coherent states}
The Walsh quantization of observables on $\t2$ uses the decomposition of the quantum
Hilbert space $\hn$ into a tensor product of ``qubits'', namely
$\hn=(\IC^3)^{\otimes k}$ (clearly, this makes sense only for $N=3^k$). 
Each discrete position $q_j=j/N$ can be represented by its ternary sequence
$q_j= 0\cdot\, \ep_0\ep_2\cdots\ep_{k-1}$, with symbols $\ep_i\in\set{0,1,2}$. 
Accordingly, each position eigenstate $\bq_j$ can be represented as a tensor product:
$$
\bq_j=e_{\ep_0}\otimes e_{\ep_2}\otimes\ldots\otimes e_{\ep_{k-1}}
=|[\ep_0\ep_2\ldots\ep_{k-1}]_k\ra\,,
$$
where $\set{e_0,\,e_1,\,e_2}$ is the canonical (orthonormal) basis of $\IC^3$. The notation
on the right hand side emphasizes the fact that this state is 
associated with the cylinder $[\bep]=[\bep]_k$
with $k$ symbols on the right of the comma, no symbol on the left. Its
image on $\t2$ is a rectangle $[\bep]_k$ of height unity and width $3^{-k}=1/N$.

Walsh quantization consists in replacing the discrete Fourier transform
\eqref{e:DFT} on $\hn$ by the Walsh(-Fourier) transform $W_{N}$, which is a unitary operator
preserving the tensor product structure of $\hn$. 
We define it through its inverse $W_{N}^*$, which maps the position basis to the orthonormal basis 
of ``Walsh momentum states'': for any $j\equiv \ep_0\ldots\ep_{k-1}$, 
$$
\bp_j= W_{N}^*\,\bq_j\defeq F^*_3 e_{\ep_{k-1}}\otimes F^*_3 e_{\ep_{k-2}}\otimes\ldots
\otimes F^*_3 e_{\ep_1}\otimes F^*_3 e_{\ep_0}
$$
(here $F_3^*$ is the inverse Fourier transform on $\IC^3$).
To agree with our notations of \S\ref{s:open-baker}, we will index the symbols relative to the
momentum coordinate by {\em negative} integers, so the Walsh momentum states will be denoted by
$$
\bp_j=|[\ep_{-k}\ldots\ep_{-1}]_0\ra\,,\quad\text{where}\quad 
p_j=j/N=0 \cdot\,\ep_{-1}\ldots \ep_{-k}\,.
$$
This momentum state is associated with a rectangle of
height $3^{-k}$ and width unity. 

More generally, for any $\ell\in\set{0,\ldots,k}$ and any sequence
$\bep=\ep_{-k+\ell}\ldots\ep_{-1}\cdot\ep_0\ldots\ep_{\ell-1}$, one can construct
a (Walsh-)coherent state
\bequ\label{e:Walsh-CS}
|[\bep]_{\ell}\ra  \defeq 
e_{\ep_0}\otimes\ldots e_{\ep_{\ell-1}}\otimes 
F_3^* e_{\ep_{-k+\ell}}\otimes\ldots F_3^* e_{\ep_{-1}} \,.
\eequ
This state is localized in the rectangle $[\bep]_\ell$ with height $3^{-k+\ell}$ and
width $3^{-\ell}$, so it still has area $1/N$ 
(all such rectangles are minimal-uncertainty, or ``quantum'' rectangles).

Like the squeezing parameter $\ssigma$ of Gaussian wavepackets (see \S\ref{s:Husimi}),
the index $\ell$ describes the aspect ratio of the coherent state. When going
to the semiclassical limit $k\to\infty$, we will
always select $\ell(k)\sim k/2$, which corresponds to an ``isotropic'' squeezing. 
One important
difference between Gaussian and Walsh coherent states lies in the fact that the latter are
{\em strictly} localized both in momentum and position. Another difference is that, for
each $\ell$,  
the $\ell$-coherent states make up a finite orthonormal basis of $\hn$, 
instead of a continuous overcomplete family.

%==========================================
\subsubsection{Walsh quantization for observables on $\Sigma$}\label{s:Lipschitz}
%==========================================

As explained in \cite{AN06}, it is more natural to Walsh-quantize observables
on the symbol space $\Sigma$ than on $\t2$.
Indeed, if one equips $\Sigma$ with the metric structure
$$
d_{\Sigma}(\bep,\bep')=\max(3^{-n_+},3^{-n_-}),\quad n_+=\min\set{n\geq 0,\,\ep_n\neq \ep'_n},\ 
n_-=\min\set{n\geq 0,\,\ep_{-n-1}\neq \ep'_{-n-1}}\,,
$$
the closed shift $\hat\sigma$ then acts as a {\it Lipschitz} map on $\Sigma$, and
the hole $\set{\ep_0=1}$ is at finite distance from its complement in $\Sigma$.
Indeed, the ``lift'' from $\t2$ to $\Sigma$ has the effect to
``blow up'' the lines of discontinuity of $B$.

The conjugacy $J:\Sigma\to\t2$ is also Lipschitz, so
any Lipschitz function $F\in Lip(\t2)$ is pushed to 
a function $f=F\circ J \in Lip(\Sigma)$. However, the converse is not true:
if we use the inverse map $(J_{|\Sigma'})^{-1}:\t2\to\Sigma'$ and take an arbitrary
$f\in Lip(\Sigma)$, the function $F= f\circ (J_{|\Sigma'})^{-1}$
is generally discontinuous on $\t2$.

Let us select some $\ell\sim k/2$. The Walsh-quantization of a function $f\in Lip(\Sigma)$ is 
defined as the following operator on $\hn$:
\bequ\label{e:Walsh-quantiz}
\Op_N^\ell(f)\defeq 3^k \sum_{[\bep]_{\ell}} 
|[\bep]_{\ell}\ra \la[\bep]_{\ell}|\;\int_{[\bep]_{\ell}}f\,d\mu_L
\eequ
Here the sum goes over all ``quantum'' cylinders $[\bep]_{\ell}$, that is
over all $3^k$ sequences $\bep=\ep_{-k+\ell}\ldots\ep_{-1}\cdot\ep_0\ldots\ep_{\ell-1}$.
The integral over $f$ is performed using the uniform Bernoulli measure $\mu_L$, which
is equivalent to the Liouville measure on $\t2$ (see the end of \S\ref{s:open-baker}).
Notice the formal similarity of this quantization with the anti-Wick 
quantization \eqref{e:anti-Wick}.

It is shown in \cite[Prop.3.1]{AN06} that this quantization satisfies the 
Axioms~\ref{d:axiom-obs} (with Lipschitz observables), and that 
two quantizations $\Op_N^{\ell_1}$, $\Op_N^{\ell_2}$ are semiclassically 
equivalent if both $\ell_1,\,\ell_2\sim k/2$.

By duality, we define the Walsh-Husimi measure of a quantum state $\psi_N$. The corresponding 
density is constant in each $\ell$-cylinder $[\bep]_{\ell}$:
\bequ\label{e:WH}
\forall \bal\in[\bep]_{\ell}\,,\qquad WH^\ell_{\psi_N}(\bal)=3^k\,|\la [\bep]_\ell , \psi_N\ra|^2\,.
\eequ
This density is originally defined on $\Sigma$, but can be 
pushed-forward to $\t2$.
Semiclassical measures
are defined as the weak-$*$ limits of sequences $(WH^\ell_{\psi_N})$,
where $N=3^k\to\infty$, and $\ell=\ell(k)\sim k/2$. One first obtains a measure $\mu^\Sigma$ 
on $\Sigma$, which can be pushed on $\t2$ into $\mu=J^*\mu^\Sigma$ (equivalently,
$\mu$ is the weak-$*$ limit of the Walsh-Husimi measures on $\t2$).

%%%%%%%%%%%%%%%%%%%%%%%%%%%%%%%%%%%%%%%%%%%%%%%%%%%%%%%%%%%%%%%%%%%%%%%
\subsection{Walsh quantization of the open baker}\label{s:Walsh-baker} 
%%%%%%%%%%%%%%%%%%%%%%%%%%%%%%%%%%%%%%%%%%%%%%%%%%%%%%%%%%%%%%%%%%%%%%%

We now recall the Walsh quantization of the open baker 
$B=B_{\vr_{sym}}$, as defined in \cite{NonZw05,NonZw06}.
Mimicking the standard quantization \eqref{e:B_N},
we replace the Fourier transforms $F_{N}^*$, $F_{N/3}$ by their Walsh analogues
$W_N$, $W_{N/3}$ (with $N=3^k$, $k\geq 0$), so that the Walsh-quantized open baker
is given by the following matrix in the position basis:
\bequ
\widetilde B_{N}\defeq W_N^*\begin{pmatrix}W_{N/3}&&\\&0&\\&&W_{N/3}\end{pmatrix}\,.
\eequ 
For any set of vectors $v_0,\,\ldots\, v_{k-1}\in \IC^3$, this operator acts as follows
on any tensor product state $v_0\otimes\ldots\otimes v_{k-1}$:
\bequ\label{e:B_k}
\widetilde B_{N}\big(v_0\otimes v_1\ldots\otimes v_{k-1}\big) =
v_1\otimes\ldots\otimes v_{k-1}\otimes \widetilde F_3^* v_0\,.
\eequ
Here $\widetilde F_3^*=F_3^*\,\pi_{02}$, where $\pi_{02}$ is the orthogonal projector on 
$\IC e_0\oplus \IC e_2$ in $\IC^3$.

One can generalize the quantum-correspondence
of \cite[Prop.3.2]{AN06} to the open shift $\sigma$, and prove the Walsh version of
Axioms~\ref{d:axiom}:
%================
\begin{prop}\label{p:Walsh-Egorov}

$i)$ Take any $f\in Lip(\Sigma)$. Then, in the limit $N=3^k\to\infty$, $\ell\sim k/2$, 
$$
\tB_N^\dagger\,\Op^\ell_N(f)\,\tB_N \sim  
\Op^{\ell}_N\big(\bbbone_{\set{\ep_0\neq 1}}\times f\circ\sigma\big)\,.
$$
Notice that the function 
$\bbbone_{\set{\ep_0\neq 1}}\times f\circ\sigma\in Lip(\Sigma)$.

$ii)$ If we take $F\in Lip_c(C(B^{-1})\sqcup \ring{H}^{-1})$ and $f=F\circ J$, we have
$$
\bbbone_{\set{\ep_0\neq 1}}\times f\circ\sigma=\big(\bbbone_{M}\times F\circ B\big)\circ J\,.
$$
\end{prop}
%================
The points $i)-ii)$ show that the family $(\tB_N)$
satisfies the Axioms~\ref{d:axiom} with respect to the map $B$ on $\t2$ and
the quantization \eqref{e:Walsh-quantiz} of observables in $Lip(\t2)$. 
The point $i)$ alone shows that $(\tB_N)$ is a quantization 
of the open shift $\sigma$ on $\Sigma$. As noticed above,
the latter interpretation allows to get rid of problems of discontinuities.
\begin{proof}
Applying \eqref{e:B_k} to a coherent state $|[\bep]_\ell\ra$, we get the exact 
evolution
$$
\tB_N^\dagger\,|[\bep]_\ell\ra=(1-\delta_{\ep_{-1},1})\,|[\sigma^{-1}(\bep)]_{\ell+1}\ra\,.
$$
That is, the coherent state $|[\bep]_\ell\ra$ is either killed if $\sigma^{-1}([\bep]_\ell)=\infty$, 
or transformed into a coherent state associated with the cylinder 
$[\sigma^{-1}(\bep)]_{\ell+1}=\sigma^{-1}([\bep]_{\ell})$.
This exact expression, which is the quantum counterpart of the classical shift \eqref{e:shift-B},
should be compared with the approximate expression \eqref{e:CS-evol}.
From there, a straighforward computation shows that, for any $f\in Lip(\Sigma)$:
$$
\tB_N^\dagger\,\Op^\ell_N(f)\,\tB_N = 
\Op^{\ell+1}_N\big((1-\delta_{\ep_{0},1})\times f\circ\sigma\big)\,.
$$
The semiclassical equivalence $\Op^{\ell+1}_N\sim \Op^{\ell}_N$
finishes the proof of $i)$.

To prove $ii)$, we remark that, 
if $F\in Lip_c(C(B^{-1})\sqcup\ring{H}^{-1})$ and $f=F\circ J$,
the function $F\circ B$ is supported away from $D(B)$, and both functions
$\bbbone_{\set{\ep_0\neq 1}}\times f\circ\sigma$ and 
$\bbbone_{\set{\ep_0\neq 1}}\times F\circ B\circ J$ are supported inside $\Sigma''$. 
The semiconjugacy \eqref{e:semiconj} shows that these two functions are equal.
Finally, we notice that, for $\bep\in\Sigma''$, one has
$\bbbone_{\set{\ep_0\neq 1}}(\bep)=\bbbone_{M}\circ J(\bep)$.
\end{proof}

%==========================================
\subsubsection{Spectrum of the Walsh open baker}
%==========================================

\begin{figure}[ht]
\rotatebox{-90}{\includegraphics[width=.5\textwidth]{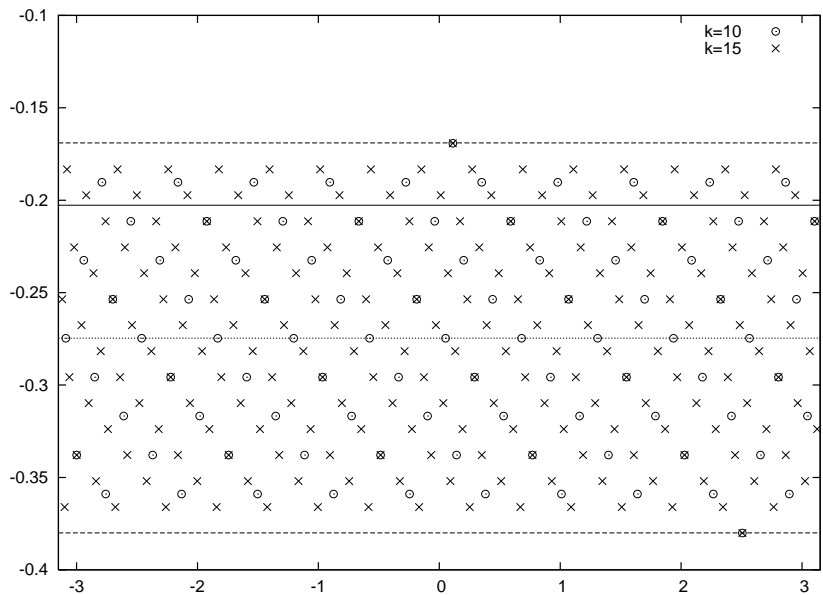}}
\caption{\label{f:toy-log} Nontrivial spectrum of the Walsh open baker 
$\tB_N$ for $N=3^{10}$ (circles) and
$3^{15}$ (crosses), using a logarithmic representation 
(horizontal=$\arg\lambda_j$, vertical=$\log|\lambda_j|$). We plot horizontal lines
at the extremal radii $|\lambda_{\pm}|$ of the spectrum (dashed),
at the radius $r_0=|\lambda_-\lambda_+|^{1/2}$ of highest degeneracies (dotted)
and at the radius corresponding to the natural measure (full).}
\end{figure}

The simple expression \eqref{e:B_k} allows to explicitly compute
the spectrum of $\tB_{N}$ (see \cite[Prop. 5.5]{NonZw06}). 
That spectrum is determined by the two nontrivial
eigenvalues $\lambda_-,\ \lambda_+$ of the matrix 
$\widetilde F_3^*$. These eigenvalues have moduli $|\lambda_+|\approx 0.8443$, 
$|\lambda_-|\approx 0.6838$.
The spectrum of $\widetilde B_{N}$ has a gap: the long-living eigenvalues are 
contained in the annulus $\set{|\lambda_-|\leq |\lambda|\leq |\lambda_+|}$,
while the rest of the spectrum lies at the origin.
Most of the eigenvalues are degenerate. If we count multiplicities, the long-living
($\equiv$ nontrivial) spectrum satisfies the following asymptotics when $k\to\infty$:
\bequ\begin{split}
\forall r>0,\qquad
&\#\set{\lambda_j\in\Spec(\tB_{N})\,,\ |\lambda_j|\geq r}=C_r\,2^k+o(2^k)\,,\\
C_r&=\begin{cases} 1\,, & r < r_0\\
0\,, & r > r_0
\end{cases}\,,\qquad r_0\defeq|\lambda_-\lambda_+|^{1/2}=3^{-1/4}\,.
\end{split}\eequ
The nontrivial
spectrum is spanned by a subspace $\cH_{N,long}$ of dimension $2^k$.
Since the trapped set for $B=B_{\vr_{sym}}$ has dimension $2d=2\,\frac{\log 2}{\log 3}$,
the above asymptotics agrees with the Fractal Weyl law \eqref{e:FWL}.
Although the density of resonances (counted with multiplicities) is 
peaked near the circle $\set{|\lambda|=r_0}$, the spectrum (as a set) 
densely fills the
annulus $\set{|\lambda_-|\leq |\lambda|\leq |\lambda_+|}$ when $k\to\infty$ 
(see Fig.~\ref{f:toy-log}). 
%%%%%%%%%%%%%%%%%%%
%%%%%%%%%%%%%%%%%%%
In the next section we construct some long-living eigenstates of $\tB_N$ and analyze
their Walsh-Husimi measures. Due to the spectral degeneracies, there is
generally a large freedom to select eigenstates $(\psi_N)$ associated with
a sequence of eigenvalues $(\lambda_N)$. Intuitively, that freedom should 
provide more possibilities for semiclassical measures.

We mention that 
J.~Keating and coworkers have recently studied the eigenstates 
of a slightly different version of the Walsh-baker, namely 
a matrix $\tB_N'$ obtained by replacing $F_3$ by the ``half-integer Fourier transform'' 
$(G_3)_{jj'}=3^{-1/2}\e^{-2\i\pi(j+1/2)(j'+1/2)/3}$ (see~\cite{KNNS07}). 

%%%%%%%%%%%%%%%%%%%%%%%%%%%%%%%%%%%%%%%%%%%%%%%%%%%%%%%%%%%%%%
\subsection{Long-living eigenstates of the Walsh open baker} 
%%%%%%%%%%%%%%%%%%%%%%%%%%%%%%%%%%%%%%%%%%%%%%%%%%%%%%%%%%%%%%%

We first provide the analogue of 
Theorem~\ref{thm:main} for the Walsh-baker. We remind that
a semiclassical measure $\mu^\Sigma$ (or its push-forward $\mu$) 
is now a weak-$*$ limit of some sequence of Walsh-Husimi measures
$(WH^{\ell}_{\psi_N})_{N=3^k\to\infty}$,
where $\psi_N$ are eigenstates of $\tB_N$, and the index $\ell\approx k/2$.

From the quantum-classical correspondence of Prop.~\ref{p:Walsh-Egorov} and
using Prop.~\ref{l:eigenmeasure-push}, we deduce the following
%%%%%%%%%%%%%%%%%%%%%%%%%
\begin{cor}\label{thm:walsh}
Let $\mu^\Sigma$ be a semiclassical measure for a sequence of long-living eigenstates 
$(\psi_N,\lambda_N)$ of 
the Walsh-baker $\tB_N$. Then:

$i)$ $\mu^\Sigma$ is an eigenmeasure for the open shift $\sigma$
on $\Sigma$, and 
the corresponding decay rate $\Lambda$ satisfies $\Lambda=\lim_{N\to\infty}|\lambda_N|^2$.

$ii)$ If $\mu^\Sigma(\Sigma\setminus\Sigma'')=0$ (where $\Sigma''$ is defined in \eqref{e:Sigma''}), 
then $\mu=J^*\mu^{\Sigma}$ is an eigenmeasure
of the open baker $B$, with decay rate $\Lambda$ .
\end{cor}
%%%%%%%%%%%%%%%%%%%%%%%%%
From the structure of $\Spec(\tB_N)$ explained above, 
there exist sequences of eigenvalues $(\lambda_N)_{N\to\infty}$ converging to any 
circle of radius $\lambda\in [|\lambda_-|,|\lambda_+|]$. We also
know that any semiclassical measure is an eigenmeasure of $\sigma$, so it is 
meaningful to ask Questions~\ref{qu1} in the present framework 
(setting $\lambda_{\max/\min}=|\lambda_{\pm}|$). We add the following question:
are there semiclassical measures $\mu^\Sigma$ such that $\mu^\Sigma(\Sigma'')<1$?
In the next section we give partial answers to these questions.

Concerning the last point in Question~\ref{qu1}, we notice that
the ``physical'' decay rate for $B=B_{\vr_{sym}}$ is $\Lambda_{nat}=2/3$. The
circle $\set{|\lambda|=\sqrt{\Lambda_{nat}}}$ is contained inside the annulus
$\set{|\lambda_-|\leq |\lambda|\leq |\lambda_+|}$ where the nontrivial spectrum
of $\tB_N$ is semiclassically dense, although it differs from
the circle $\set{|\lambda|=r_0}$ where the spectral density is peaked, 
see Fig.~\ref{f:toy-log}. 
Still, there exist semiclassical measures of $\tB_N$ with
eigenvalue $\Lambda_{nat}$, and it is relevant to ask whether $\mu_{nat}$ 
can be one of these. 
At present we are not able to answer that question. The next section
shows that there are plenty of semiclassical measures with eigenvalue $\Lambda_{nat}$,
so even if $\mu_{nat}$ is a semiclassical measure with $\Lambda=\Lambda_{nat}$, 
it is certainly not the only one.

%%%%%%%%%%%%%%%%%%%%%%%%%%%%%%%%%%%%%%%%%%%%%%%%%%%%%%%%%%%%
\subsection{Constructing the eigenstates of $\tB_N$}
%%%%%%%%%%%%%%%%%%%%%%%%%%%%%%%%%%%%%%%%%%%%%%%%%%%%%%%%%%%%

In this section we construct one particular (right)
eigenbasis of $\tB_N$ restricted to the subspace $\cH_{N,long}$ of long-living
eigenstates.
The construction starts from the (right) eigenvectors
$v_\pm\in\IC^3$
of $\widetilde F_3^*$ associated with $\lambda_{\pm}$. 
Notice that these two vectors (which we take normalized)
are not orthogonal to each other.
For any sequence 
$\bet=\eta_0\ldots\eta_{k-1}$, $\eta_i\in\set{\pm}$, we form the tensor product state 
$$
|\bet\ra\defeq v_{\eta_0}\otimes v_{\eta_1}\ldots\otimes v_{\eta_{k-1}}\,.
$$
The action \eqref{e:B_k} of $\tB_N$ implies that
$$
\tB_N |\bet\ra = \lambda_{\eta_0}\,|\tau(\bet)\ra\,,
$$
where $\tau$ acts as a cyclic shift on the sequence: 
$\tau(\eta_0\ldots\eta_{k-1})=\eta_1\ldots\eta_{k-1}\eta_0$.
The {\em orbit} $\set{\tau^j(\bet),\ j\in\IZ}$ contains 
$\ell_{\bet}$ elements, where the {\em period} $\ell_{\bet}$ of the 
sequence $\bet$ necessarily divides $k$.
The states $\set{|\tau^j(\bet)\ra,\ j=0,\ldots,\ell_{\bet}-1}$ are not orthogonal 
to each other, but form a linearly 
independent family, which generates the $\tB_N$-invariant subspace 
$\cH_{\bet}\subset \cH_{N,long}$. The eigenvalues
of $\tB_N$ restricted to $\cH_{\bet}$ are of the form 
$\lambda_{\bet,r}=\e^{2\i\pi r/\ell_{\bet}}\,
\big(\prod_{j=0}^{\ell_{\bet}-1}\lambda_{\eta_j}\big)^{1/\ell_{\bet}}$ 
(with indices $r=1,\ldots,\ell_{\bet}$), 
and the corresponding eigenstates read
\bequ\label{e:formula}
|\psi_{\bet,r}\ra=\frac1{\sqrt{\cN_{\bet,r}}}
\sum_{j=0}^{\ell_{\bet}-1} c_{\bet,r,j}\; |\tau^j(\bet)\ra\,,\qquad 
c_{\bet,r,j}=\prod_{m=0}^{j-1}\frac{\lambda_{\eta_m}}
{\lambda_{\bet,r}}\,,
\eequ
where $\cN_{\bet,r}>0$ is the factor which normalizes $|\psi_{\bet,r}\ra$.
Up to a phase, this state is unchanged if $\bet$ is replaced by $\tau(\bet)$.
In the next subsections 
we explicitly compute the Walsh-Husimi measures of some of these eigenstates.
%%%%%%%%%%%%%%%%%%%
\begin{figure}[htbp]
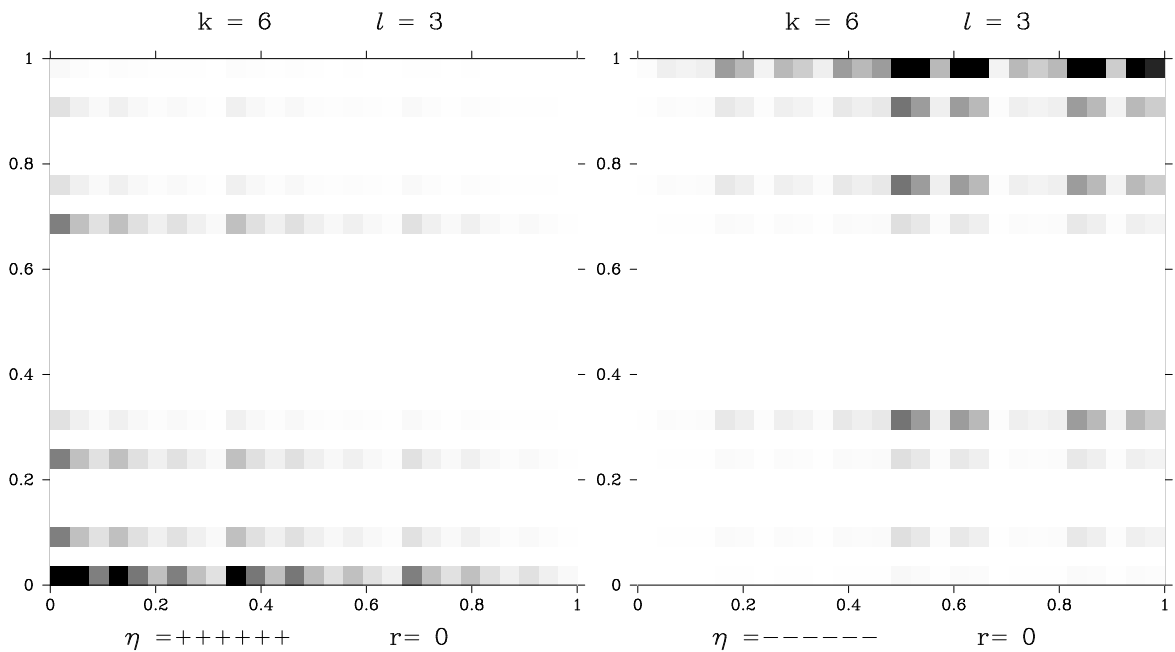

\begin{center}
\rotatebox{-90}{\includegraphics[width=8.5cm]{max6.ps}}
\rotatebox{-90}{\includegraphics[width=8.5cm]{min6.ps}}
\caption{\label{f:maximal} Walsh-Husimi densities for the extremal eigenstates $\psi_{+,N}$,
$\psi_{-,N}$ of $\tB_{N}$, with $N=3^6$, $\ell=3$. These are coarse-grained 
versions of the Bernoulli measures $\mu_{\vP_{+}}$ and $\mu_{\vP_{-}}$.}
\end{center}
\end{figure}
%%%%%%%%%%%%%%%%%%%

%%%%%%%%%%%%%%%%%%%%%%%%%%%%%%%%%%%%%%%%%%%%%%%%%%%%%%%%%%%%%%%%%%%%%%%
\subsubsection{Extremal eigenstates} \label{s:extremal}
%%%%%%%%%%%%%%%%%%%%%%%%%%%%%%%%%%%%%%%%%%%%%%%%%%%%%%%%%%%%%%%%%%%%%%%

The simplest case is provided by the sequence $\bet=++\cdots+$, which has period $1$, so that
$|\psi_{+,N}\ra=|\bet\ra=v_+^{\otimes k}$ is the (unique) eigenstate 
associated with the
largest eigenvalue $\lambda_+$ (this is the longest-living eigenstate).
For any choice of index 
$0\leq \ell\leq k$, the Walsh-Husimi measure of $|\psi_{+,N}\ra$ factorizes:
$$
\text{for all $\ell$-rectangle }[\bep]_\ell,\qquad
WH^\ell_{\psi_{+,N}}([\bep]_\ell)
=\prod_{j=0}^{\ell-1} |\la v_{+},\,e_{\ep_j}\ra|^2\ 
\prod_{j=-1}^{-k+\ell}|\la v_{+},\,F_3^*\, e_{\ep_j}\ra|^2\,.
$$
The second product involves the vector $w_+\defeq F_3 v_+$, with components
$w_{+,\ep}=(1-\delta_{\ep,1})v_{+,\ep}/\lambda_+$.
Following the notations of \S\ref{s:Bernoulli}, let $\mu^\Sigma_{\vP_+}$ 
be the Bernoulli eigenmeasure of $\sigma$ with weights 
$P_{+,\ep}=|v_{+,\ep}|^2$, $P^*_{+,\ep}=|w_{+,\ep}|^2$.
The above expression shows that the Husimi measure $WH^\ell_{\psi_{+,N}}$ is equal
to the measure  $\mu^\Sigma_{\vP_+}$, 
{\em conditioned on the grid formed by the $\ell$-cylinders}. 
Since the diameters of the cylinders decrease to
zero as $k\to\infty$, $\ell(k)\sim k/2$, the Husimi measures 
$(H^\ell_{\psi_{+,N}})$ converge to $\mu^\Sigma_{\vP_+}$.

One can similarly show that the Husimi functions of the eigenstates
$\psi_-=v_-^{\otimes k}$, associated with the smallest nontrivial eigenvalue $\lambda_-$, 
converge to the Bernoulli
eigenmeasure $\mu^\Sigma_{\vP_-}$, with weights $P_{-,\ep}=|v_{-,\ep}|^2$, 
$P^*_{-,\ep}=|w_{-,\ep}|^2$, where $w_{-}=F_3 v_{-}$.

In Fig.~\ref{f:maximal} we plot the Walsh-Husimi densities (pushed-forward on $\t2$)
for $\psi_{+,N}$ and $\psi_{-,N}$,
using the ``isotropic'' $\ell=k/2$. These give a clear idea of the
selfsimilar structure of the respective semiclassical measures 
$\mu_{\vP_+}$ and $\mu_{\vP_-}$. The weights have the approximate values
$\vP_+\approx (0.579, 0.287, 0.134)$,
$\vP_-\approx (0.088, 0.532, 0.380)$.

Considering the fact that the eigenvalues $\lambda_N$ close to the circles 
of radii $|\lambda_+|$
and $|\lambda_-|$ have small degeneracies, we propose the following
%&&&&&&&&&&&&&
\begin{conj}
Any sequence of eigenstates $(\psi_N)_{N\to\infty}$ with eigenvalues
$|\lambda_N|\to |\lambda_+|$ (resp.
$|\lambda_N|\to |\lambda_-|$) converges
to the semiclassical measure $\mu^\Sigma_{\vP_+}$ (resp. $\mu^\Sigma_{\vP_-}$).
\end{conj}
%&&&&&&&&&&&&&
This conjecture can be proven for the version of the Walsh baker 
$\tB_N'$ studied in \cite{KNNS07}: in 
that case the two eigenvectors of $\tilde G^*_3$ replacing $v_{\pm}$
are orthogonal to each other, which greatly simplifies the analysis.
The limit measure $\mu^\Sigma_{\vP'_+}$ is then the ``uniform'' measure 
on the trapped set $\set{\ep_n\neq 1,\ n\in\IZ}$, 
with $\vP'_+={\vP'_+}^*=(1/2,0,1/2)$.

%%%%%%%%%%%%%%%%%%%%%%%%%%%%%%%%%%%%%%%%%%%%%%%%%%%%%%%%%%%%%%%%%%%%%%%
\subsubsection{Semiclassical measures in the ``bulk''} 
%%%%%%%%%%%%%%%%%%%%%%%%%%%%%%%%%%%%%%%%%%%%%%%%%%%%%%%%%%%%%%%%%%%%%%%

%%%%%%%%%%%%%%%%%%%
\begin{figure}[htbp]
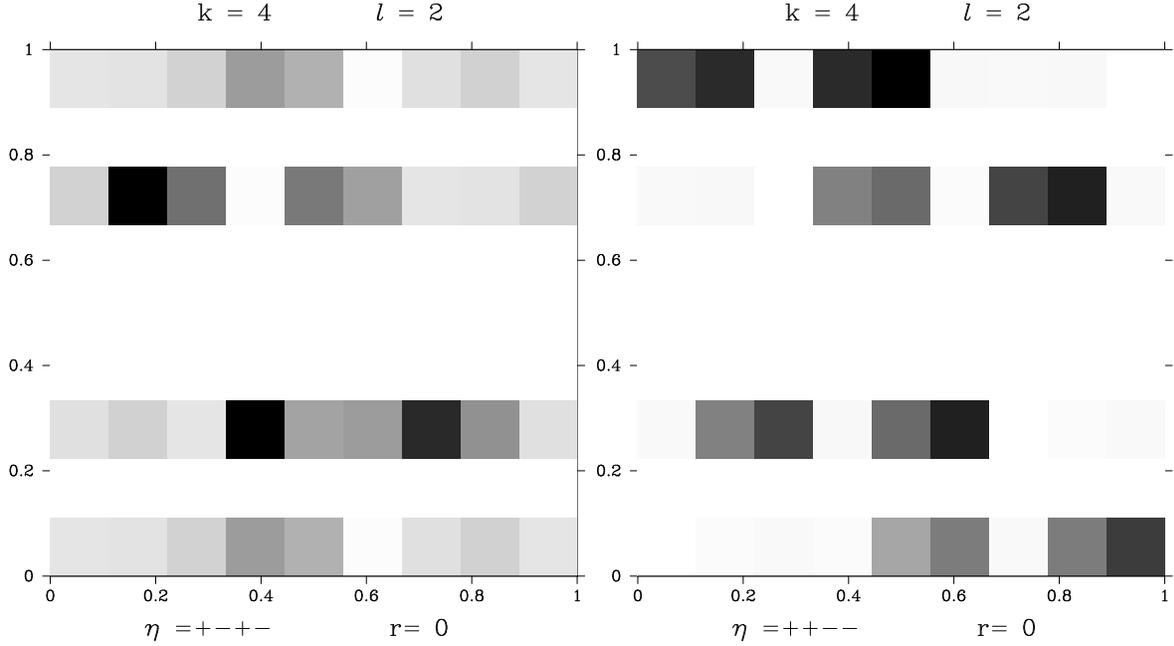

\begin{center}
\rotatebox{-90}{\includegraphics[width=8.5cm]{mix+-+-4-0.ps}}
\rotatebox{-90}{\includegraphics[width=8.5cm]{mix++--4-0.ps}}
\caption{\label{f:mix} Walsh-Husimi densities of two eigenstates $\psi_{\bet,0}$
constructed from the sequences $\bet_{4}=+-+-$ (left) and $\bet'_{4}=++--$ (right).}
\end{center}
\end{figure}
%%%%%%%%%%%%%%%%%%%

In this section we investigate some eigenstates of the form \eqref{e:formula},
with eigenvalues $\lambda_N$ situated in the ``bulk'' of the nontrivial spectrum, 
that is $|\lambda_N|\in (|\lambda_-|,|\lambda_+|)$. 

In the next proposition we show that there can be two different semiclassical
measures with the same decay rate $\Lambda$. 
This answers by the {\em negative} the first point in Question~\ref{qu1}.
%%%%%%%%%%%%%
\begin{prop}
Choose a rational number $t=\frac{m}{n}$, with $n\geq 1$, $1\leq m\leq n-1$.

i) Select a sequence $\bet_n$ with $m$ $(+)$ and $n-m$ $(-)$. 
For all $k'\geq 1$, form the repeated sequence $(\bet_n)^{k'}$, and
choose $r=r(k')\in\IZ$
arbitrarily. Then the sequence of eigenstates 
$\big(\psi_{(\bet_n)^{k'},r(k')}\big)_{k=nk'\to\infty}$ 
converges to the semiclassical measure $\mu^\Sigma_{\bet_n}$, which is
a linear combination of Bernoulli measures for the iterated shift $\sigma^n$ (see \eqref{e:rho_bet_n}). 
This measure is independent of the choice of $\big(r(k')\big)$, and has decay rate 
$\Lambda_t\defeq\big|\lambda_-^{1-t}\,\lambda_+^{t}\big|^2$. Its push-forward is an
eigenmeasure of $B$.

ii) If $\bet_n$
and $\bet'_n$ are two sequences with $m$ $(+)$ and $n-m$ $(-)$, 
which are not related by a cyclic permutation, then 
the semiclassical measures $\mu^\Sigma_{\bet_n}$,
$\mu^\Sigma_{\bet'_{n}}$ are mutually singular, and so are their
push-forwards on $\t2$.
\end{prop}
%%%%%%%%%%%%%
Note that the radius $\sqrt{\Lambda_t}$ lies in the ``bulk'' of the 
nontrivial spectrum, but can be different from the radius $r_0$ where the spectral 
density is peaked.
In Fig.~\ref{f:mix} we plot the Husimi functions of two states $\psi_{\bet_4,0}$,
$\psi_{\bet'_4,0}$
constructed from two $4$-sequences $\bet_{4}$, $\bet'_{4}$ not cyclically related. 
The two functions, which give a rough idea of the limit measures $\mu_{\bet_{4}}$, 
$\mu_{\bet'_{4}}$, seemingly concentrate on different parts of the phase space.
\begin{proof}
For short we call $\bet=(\bet_n)^{k'}$ which has period $\ell_{\bet}=\ell_{\bet_n}$, and $r=r(k')$. 
From \eqref{e:formula}, each state $|\psi_{\bet,r}\ra$ is a combination of 
$\ell_{\bet}$ states $|\tau^j(\bet)\ra$, and its eigenvalue has exact modulus $\sqrt{\Lambda_t}$.
When $k'\to\infty$, those states are asymptotically orthogonal to
each other. Indeed, their overlaps can be decomposed as
$$
\la \bet|\tau^j(\bet)\ra=\big(\la \bet_n|\tau^j(\bet_n)\ra \big)^{k'}\,
\quad j=0,\ldots,\ell_{\bet}-1\,,
$$
and for any $j\neq 0\bmod \ell_{\bet}$ we have $\bet_n\neq \tau^j(\bet_n)$, which implies
$|\la \bet_n|\tau^j(\bet_n)\ra|\leq c$, with $c \defeq |\la v_+,v_-\ra|^2<1$.
As a result, the normalization factor of $\psi_{\bet,r}$ satisfies
$$
\cN_{\bet,r}= \sum_{j=0}^{\ell_{\bet}-1} |c_{\bet,r,j}|^2+\cO(c^{k'})\,,\qquad k'\to\infty\,.
$$
To study the semiclassical measures of the sequence $(\psi_{\bet,r})_{k'\to\infty}$, 
we fix some cylinder
$[\bal]=[\alpha_{-l'}\ldots\alpha_{-1}\cdot \alpha_0\ldots\alpha_{l-1}]$
and compute the weight of the measures $H^\ell_{\psi_{\bet,r}}$. If $k$ is large enough and
$\ell\sim k/2$, the conditions $\ell>l$ and $k-\ell>l'$ are fulfilled, and this weight
can be written
\bequ\label{e:overlap}
H^\ell_{\psi_{\bet,r}}([\bal])=
\la\psi_{\bet,r}|\Pi_{[\bal]}|\psi_{\bet,r}\ra=
\sum_{j,j'=0}^{\ell_{\bet}-1} c_{\bet,r,j}\,\bar{c}_{\bet,r,j'}\,
\la\tau^{j'}(\bet)|\Pi_{[\bal]}|\tau^j(\bet)\ra\,,
\eequ
where the projector on $[\bal]$ is a tensor product operator:
$$
\Pi_{[\bal]}=\pi_{\alpha_0}\otimes\pi_{\alpha_1}\ldots\pi_{\alpha_{l-1}}\otimes 
(I)^{\otimes k-l-l'}\otimes F_3^*\pi_{\alpha_{-l'}}F_3\otimes\ldots \otimes F_3^*\pi_{\alpha_{-1}}F_3\,.
$$
The tensor factor $(I)^{k-l-l'-1}$ implies that each matrix element 
$\la\tau^{j'}(\bet)|\Pi_{[\bal]}|\tau^j(\bet)\ra$ contains
a factor $\big(\la\tau^{j'}(\bet_n)|\tau^j(\bet_n)\ra\big)^{k'-\cO(1)}$; 
for the same reasons as above, this element
is $\cO(c^{k'})$ if $j\neq j'$. We are then lead to consider only
the diagonal elements $j=j'$:
\bequ\label{e:weight}
\forall j=0,\ldots,\ell_{\bet}-1,\qquad
\la\tau^{j}(\bet)|\Pi_{[\bal]}|\tau^j(\bet)\ra=
\prod_{i=0}^{l-1} P_{\eta_{j+i},\alpha_i}\,
\prod_{i'=1}^{l'} P^*_{\eta_{j-i'},\alpha_{-i'}}\,.
\eequ
As above the weights
$P_{\pm,\ep}=|v_{\pm,\ep}|^2$, $P^*_{\pm,\ep}=|w_{\pm,\ep}|^2$,
and the definition of $\eta_i$ was extended to $i\in\IZ$ by periodicity. We claim that
the right hand side exactly corresponds to $\tilde\mu^\Sigma_{\tau^j(\bet_n)}([\bal])$,
where $\tilde\mu^\Sigma_{\tau^j(\bet_n)}$ is a certain Bernoulli eigenmeasure for the iterated
shift $\sigma^n$.
The latter can be seen as a simple shift on the symbol space $\tilde\Sigma$ constructed from
$3^n$ symbols $\teps\in\set{0,\ldots,3^n-1}$: 
each $\teps$ is in one-to-one correspondence
with a certain $n$-sequence $\ep_0\ldots\ep_{n-1}$.
Adapting the formalism of \S\ref{s:Bernoulli}
to this new symbol space,
the Bernoulli measure $\tilde\mu^\Sigma_{\tau^j(\bet_n)}$ corresponds to the
following weight distributions $\tilde\vP$, $\tilde\vP^*$:
\bequ\label{e:weights}
\text{for}\ \tilde\eps\equiv \ep_0\ldots\ep_{n-1},\qquad
\tilde P_{\tilde\ep}=\prod_{i=0}^{n-1} P_{\eta_{n,j+i},\ep_i}\quad\text{and}\quad
\tilde P^*_{\tilde\ep}=\prod_{i=1}^{n} P^*_{\eta_{n,j-i},\ep_{n-i}}\,.
\eequ
The measures $\tilde\mu^\Sigma_{\tau^j(\bet_n)}$, $j=0,\ldots,\ell_{\bet}-1$,
are related to one another through $\sigma$:
\bequ\label{e:class-cycle}
\sigma^*\,\tilde\mu^\Sigma_{\tau^j(\bet_n)}= |\lambda_{\eta_{n,j}}|^2\,
\tilde\mu^\Sigma_{\tau^{j+1}(\bet_n)}\,.
\eequ
Finally, the semiclassical measure associated with the sequence
$\big(\psi_{\bet,r}\big)_{k=nk'\to\infty}$ is
\bequ\label{e:rho_bet_n}
\mu^\Sigma_{\bet_n}=
\frac1{\cN_{\bet_n}}\sum_{j=0}^{\ell_{\bet}-1} C_{\bet_n,j}\,\tilde\mu^\Sigma_{\tau^j(\bet_n)}\,,\quad
\text{where}\quad C_{\bet_n,j}=\prod_{m=0}^{j-1}\frac{|\lambda_{\eta_{n,m}}|^2}{\Lambda_t}\,,
\quad \cN_{\bet_n}=\sum_{j=0}^{n-1}C_{\bet_n,j}
\eequ
This is
a probability eigenmeasure of $\sigma$, with decay rate $\Lambda_t$. It only 
depends on the orbit $\set{\tau^j\bet_n,\ j=0,\ldots,\ell_{\bet_n}-1}$, and not on the
choice of $(r(k'))$. This measure does not charge $\Sigma\setminus\Sigma''$, so its push-forward is an
eigenmeasure of $B$.

The proof of statement $ii)$ goes as follows: since $\bet_n$ and $\bet'_n$ are
not cyclically related, for
any $j,j'\in\IZ$, the weight distributions $\tilde\vP$ and $\tilde\vP'$ 
defining respectively the 
Bernoulli measures $\tilde\mu^\Sigma_{\tau^j(\bet_n)}$ and
$\tilde\mu^\Sigma_{\tau^{j'}(\bet'_n)}$ (see \eqref{e:weights}) are different. 
As a result, these two measures are mutually singular
(see the end of \S\ref{s:Bernoulli}), and so are the 
two linear combinations
$\mu^\Sigma_{\bet_n}$, $\mu^\Sigma_{\bet'_n}$. Because the weights $\tilde\vP$, $\tilde\vP'$
are different from $(1,0,\ldots,0)$ and $(0,\ldots,1)$, the push-forwards $\mu_{\bet_n}$,
$\mu_{\bet'_n}$ are also mutually singular.
\end{proof}
By a standard density argument, we can exhibit semiclassical measures for arbitrary decay
rates in $[|\lambda_-|^2,|\lambda_+|^2]$.
\begin{cor}\label{c:eigenmeasures}
Consider a real number $t\in [0,1]$, and a 
sequence of rationals $(t_p = \frac{m_p}{n_p}\in [0,1]\big)_{p\to\infty}$ converging
to $t$ when $p\to\infty$. 
For each $p$, let $\bet_p$ be a sequence with $m_p$
$(+)$ and $n_p-m_p$ $(-)$, and $\mu^\Sigma_{\bet_p}$ the 
$\sigma$-eigenmeasure constructed above.
Then any weak-$*$ limit of the sequence
$\big(\mu^\Sigma_{\bet_p}\big)_{p\to\infty}$ is a semiclassical measure of $(\tB_N)$;
it is a $\sigma$-eigenmeasure  of decay rate 
$\Lambda_{t}=\big|\lambda_-^{1-t}\,\lambda_+^{t}\big|^2$, and
its push-forward is an eigenmeasure of $B$. 
\end{cor}
Let us call $\mathfrak{M}^\Sigma(\Lambda_t)$ the family of semiclassical measures obtained
this way, and 
$\mathfrak{M}^\Sigma=\cup_{t\in[0,1]}\mathfrak{M}^\Sigma(\Lambda_t)$.
We don't know whether the family $\mathfrak{M}^\Sigma$
exhausts the full set of semiclassical measures for $(\tB_N)$. Still, 
we can address the third point in Question~\ref{qu1} with respect to this family.
\begin{prop}
The family of eigenmeasures $\mathfrak{M}^\Sigma$ does not contain
the natural measure $\mu^\Sigma_{nat}=\mu^\Sigma_{\vr_{sym}}$. The same statement
holds after push-forward on $\t2$.
\end{prop}
\begin{proof}
For any $\ep_0\in\set{0,1,2}$, let us compute the weight of a
measure $\mu^\Sigma_{\bet_p}\in\mathfrak{M}^\Sigma$ on the cylinder
$[\eps_0]$ (corresponding to the vertical rectangle $\overline{R_{\ep_0}}$). For the 
natural measure we have
$\mu^\Sigma_{nat}([\ep_0])=1/3$ for $\ep_0=0,1,2$.

From \eqref{e:weights}, for any $j\in\IZ$ one has
$\tilde\mu^\Sigma_{\tau^j(\bet_p)}([\eps_0])= P_{\eta_{p,j},\ep_0}$. Combining this
with \eqref{e:rho_bet_n}, we get
$$
\mu^\Sigma_{\bet_p}([\ep_0])=C_{+}\,P_{+,\ep_0}+ (1-C_+)\,P_{-,\ep_0},\quad\text{where}\quad
C_{+}=\frac1{\cN_{\bet_p}}\sum_{j\,:\,\bet_{p,j}=(+)} C_{\bet_p,j}\,.
$$
For any $\mu^\Sigma\in\mathfrak{M}$, the weights $\mu^\Sigma([\ep_0])$  will take 
the same form, for some $C_{+}\in [0,1]$. Using the approximate expressions for
the weights given in \S\ref{s:extremal},
one checks that the condition 
$\mu^\Sigma([\ep_0])=1/3$ cannot be satisfied simultaneously for $\ep_0=0,1,2$.
\end{proof}
%%%%%%%%%%%%%%%%%%%%%%%%%%%%%%%%%%%%%%%%%%%%%%%%%%%%%%%%%%%%%%%%%%%%%%%
%%%%%%%%%%%%%%%%%%%%%%%%%%%%%%%%%%%%%%%%%%%%%%%%%%%%%%%%%%%%%%%%%%%%%%%
\section{Concluding remarks}
%%%%%%%%%%%%%%%%%%%%%%%%%%%%%%%%%%%%%%%%%%%%%%%%%%%%%%%%%%%%%%%%%%%%%%%
%%%%%%%%%%%%%%%%%%%%%%%%%%%%%%%%%%%%%%%%%%%%%%%%%%%%%%%%%%%%%%%%%%%%%%%

The main result of this paper is the semiclassical connection between, on the one hand,
eigenfunctions of a quantum open map (which mimick ``resonance eigenfunctions''), on the
other hand, eigenmeasures of the classical open map. 

We proved that, modulo some problems at the discontinuities of the classical map, 
semiclassical measures 
associated with long-living resonant:
are eigenmeasures of
the classical dynamics, and their decay rate is directly related with 
those of the corresponding resonant eigenstates (see Thm~\ref{thm:main}).
This result, which basically derives from Egorov's theorem, has been expressed
in a quite general framework, and applied to the specific example of the open baker's map. 
An analogue has been proven
for the more realistic setting of Hamiltonian scattering \cite[Theorem~3]{NonZw-gap}.

Although the construction and classification of eigenmeasures with decay rates $\Lambda<1$
is quite easy, the classification of semiclassical measures
among all possible eigenmeasures remains largely open (see Question~\ref{qu1}).
The solvable model provided by the
Walsh-quantized baker provides some hints to these classification, in the form
of an explicit family of semiclassical measures,
but we have no idea whether these results
apply to more general systems, not even the ``standard'' quantum open baker. Indeed,
the high degeneracies of the Walsh-baker may be responsible for a nongeneric profusion 
of semiclassical measures for that model.
Interestingly, the natural eigenmeasure does not seem to play
a particular role at the quantum level. 

A tempting way of constraining the set of semiclassical
measures would be to adapt the ``entropic'' methods of \cite{AN06} to open chaotic maps. 
A desirable output of these methods would be, for instance, to forbid
semiclassical measures from being of the pure point type described in \S\ref{s:pp}.

%%%%%%%%%%%%%%%%%%%%%%%%%%%%%%%%%%%%%%%%%%%%%%%%%%%%%%%%%%%%%%%%%%%%%%%
%%%%%%%%%%%%%%%%%%%%%%%%%%%%%%%%%%%%%%%%%%%%%%%%%%%%%%%%%%%%%%%%%%%%%%%

\end{document}

%% file: def.tex
\newcommand{\nwc}{\newcommand}
\nwc{\nwt}{\newtheorem}
\nwt{coro}{Corollary}

%font change

\nwc{\mf}{\mathbf} %Latex (as in \bf not tilted math letters)
\nwc{\blds}{\boldsymbol} %Latex 
\nwc{\ml}{\mathcal} %Latex

%greek letters

\nwc{\lam}{\lambda}
\nwc{\del}{\delta}
\nwc{\Del}{\Delta}
\nwc{\Lam}{\Lambda}
\nwc{\elll}{\ell}
%blackboard bold math

\nwc{\IA}{\mathbb{A}} %algebraic
\nwc{\IB}{\mathbb{B}} %ball
\nwc{\IC}{\mathbb{C}} %complex
\nwc{\ID}{\mathbb{D}} %Dedekind
\nwc{\IE}{\mathbb{E}} %Euklides
\nwc{\IF}{\mathbb{F}} %finite field
\nwc{\IG}{\mathbb{G}} %Gauss
\nwc{\IH}{\mathbb{H}} %Hilbert\N-subgroup
\nwc{\IN}{\mathbb{N}} %natural
\nwc{\IP}{\mathbb{P}} %prime
\nwc{\IQ}{\mathbb{Q}} %rational
\nwc{\IR}{\mathbb{R}} %real
\nwc{\IS}{\mathbb{S}} %sphere
\nwc{\IT}{\mathbb{T}} %torus
\nwc{\IZ}{\mathbb{Z}} %integers
\def\bbbone{{\mathchoice {1\mskip-4mu {\rm{l}}} {1\mskip-4mu {\rm{l}}}
{ 1\mskip-4.5mu {\rm{l}}} { 1\mskip-5mu {\rm{l}}}}}

%Straight (vector) bold letters

%lowercase

\nwc{\va}{{\bf a}}
\nwc{\vb}{{\bf b}}
\nwc{\vc}{{\bf c}}
\nwc{\vd}{{\bf d}}
\nwc{\ve}{{\bf e}}
\nwc{\vf}{{\bf f}}
\nwc{\vg}{{\bf g}}
\nwc{\vh}{{\bf h}}
\nwc{\vi}{{\bf i}}
\nwc{\vj}{{\bf j}}
\nwc{\vk}{{\bf k}}
\nwc{\vl}{{\bf l}}
\nwc{\vm}{{\bf m}}
\nwc{\vn}{{\bf n}}
\nwc{\vo}{{\it o}}
\nwc{\vp}{{\bf p}}
\nwc{\vP}{{\bf P}}
\nwc{\vq}{{\bf q}}
\nwc{\vr}{{\bf r}}
\nwc{\vs}{{\bf s}}
\nwc{\vt}{{\bf t}}
\nwc{\vu}{{\bf u}}
\nwc{\vv}{{\bf v}}
\nwc{\vw}{{\bf w}}
\nwc{\vx}{{\bf x}}
\nwc{\vy}{{\bf y}}
\nwc{\vz}{{\bf z}}

%bold letters
%\b* letters are tilted in math mode and scale in equations. 
%but cannot be used in plain text format.

%I. lowercase

\nwc{\bk}{\blds{k}}
\def\k{\blds{k}}
\nwc{\bm}{\blds{m}}
\nwc{\bp}{\blds{p}}
\nwc{\bq}{\blds{q}}
\nwc{\bn}{\blds{n}}
\nwc{\bv}{\blds{v}}
\nwc{\bw}{\blds{w}}
\nwc{\bx}{x}
%\nwc{\bx}{\blds{x}}
\nwc{\bxi}{\blds{\xi}}
\nwc{\by}{\blds{y}}
\nwc{\bz}{\blds{z}}

\nwc{\bal}{\blds{\alpha}}
\nwc{\bep}{\blds{\epsilon}}
\nwc{\bnu}{\blds{\nu}}
\nwc{\bmu}{\blds{\mu}}
\nwc{\bet}{\blds{\eta}}
%caligraphic

\nwc{\cA}{\ml{A}}
\nwc{\cB}{\ml{B}}
\nwc{\cC}{\ml{C}}
\nwc{\cD}{\ml{D}}
\nwc{\cE}{\ml{E}}
\nwc{\cF}{\ml{F}}
\nwc{\cG}{\ml{G}}
\nwc{\cH}{\ml{H}}
\nwc{\cI}{\ml{I}}
\nwc{\cJ}{\ml{J}}
\nwc{\cK}{\ml{K}}
\nwc{\cL}{\ml{L}}
\nwc{\cM}{\ml{M}}
\nwc{\cN}{\ml{N}}
\nwc{\cO}{\ml{O}}
\nwc{\cP}{\ml{P}}
\nwc{\cQ}{\ml{Q}}
\nwc{\cR}{\ml{R}}
\nwc{\cT}{\ml{T}}
\nwc{\cU}{\ml{U}}
\nwc{\cV}{\ml{V}}
\nwc{\cW}{\ml{W}}
\nwc{\cX}{\ml{X}}
\nwc{\cY}{\ml{Y}}
\nwc{\cZ}{\ml{Z}}

% tilde

\nwc{\ctGamma}{\complement\tilde\Gamma}
\nwc{\tGamma}{\tilde\Gamma}
\nwc{\tA}{\widetilde{A}}
\nwc{\tB}{\widetilde{B}}
\nwc{\tR}{\tilde{R}}
\nwc{\tmu}{\widetilde{\mu}}
\nwc{\teps}{\tilde\epsilon}

%\nwc{\cS}{\ml{S}}
\nwc{\cS}{M}

\nwc{\hM}{\hat{M}}
\nwc{\hT}{\hat{T}}
\nwc{\hB}{\hat{B}}

%miscellany
\nwc{\To}{\longrightarrow} %limits

\nwc{\ad}{\rm ad}
\nwc{\eps}{\epsilon}
\nwc{\ep}{\epsilon}
\nwc{\vareps}{\varepsilon}
\nwc{\ssigma}{s}

\def\ep{\epsilon}
\def\tr{{\rm tr}}
\def\Tr{{\rm Tr}}
\def\i{{\rm i}}
\def\mi{{\rm i}}
\def\e{{\rm e}}
\def\sq2{\sqrt{2}}
\def\sqn{\sqrt{N}}
\def\vol{\mathrm{vol}}
\def\defi{\stackrel{\rm def}{=}}
\def\t2{{\mathbb T}^2}
\def\tt2{{\mathbb T}^2}
%\nwc{\t1}{{\mathbb T}^1}
\def\s2{{\mathbb S}^2}
\def\hn{\mathcal{H}_{N}}
\def\shbar{\sqrt{\hbar}}
\def\A{\mathcal{A}}
\def\N{\mathbb{N}}
\def\T{\mathbb{T}}
\def\R{\mathbb{R}}
\def\Z{\mathbb{Z}}
\def\C{\mathbb{C}}
\def\O{\mathcal{O}}
\def\Sp{\mathcal{S}_+}
\nwc{\rest}{\restriction}
\nwc{\diam}{\operatorname{diam}}
\nwc{\Res}{\operatorname{Res}}
\nwc{\Spec}{\operatorname{Spec}}
\nwc{\Vol}{\operatorname{Vol}}
\nwc{\Op}{\operatorname{Op}}
\nwc{\supp}{\operatorname{supp}}
\nwc{\Span}{\operatorname{span}}
\nwc{\OpAW}{{\rm Op}^{{\rm AW}\!,\ssigma}_N}

\def\hto0{\xrightarrow{h\to 0}}
\def\Nto8{\xrightarrow{N\to \infty}}
\def\htoo{\stackrel{h\to 0}{\longrightarrow}}
\def\rto0{\xrightarrow{r\to 0}}

\providecommand{\abs}[1]{\lvert#1\rvert}
\providecommand{\norm}[1]{\lVert#1\rVert}
\providecommand{\set}[1]{\left\{#1\right\}}
\providecommand{\integ}[1]{\lfloor#1\rfloor}

\nwc{\la}{\langle}
\nwc{\ra}{\rangle}
\nwc{\lp}{\left(}
\nwc{\rp}{\right)}

%\nwc{\bal}{\begin{align}}
\nwc{\bequ}{\begin{equation}}
\nwc{\ben}{\begin{equation*}}
\nwc{\bea}{\begin{eqnarray}}
\nwc{\bean}{\begin{eqnarray*}}
\nwc{\bit}{\begin{itemize}}
\nwc{\bver}{\begin{verbatim}}

%\nwc{\eal}{\end{align}}
\nwc{\eequ}{\end{equation}}
\nwc{\een}{\end{equation*}}
\nwc{\eea}{\end{eqnarray}}
\nwc{\eean}{\end{eqnarray*}}
\nwc{\eit}{\end{itemize}}
\nwc{\ever}{\end{verbatim}}

\newcommand{\defeq}{\stackrel{\rm{def}}{=}}

%% file: math7.bbl
\begin{thebibliography}{999999}

\bibitem{AN06}
N.~Anantharaman and S.~Nonnenmacher, 
{\em Entropy of semiclassical measures of the Walsh-quantized baker's map}, 
Ann. Henri Poincar\'e {\bf 8} (2007) 37--74

\bibitem{BV89}
N.~L.~Balazs and A.~Voros, {\em The quantized baker's transformation},
Ann. Phys. (NY) {\bf 190} (1989) 1--31

\bibitem{BU03}
D.~Borthwick and A.~Uribe, {\em On the pseudospectra of Berezin-Toeplitz operators}, 
Meth. Appl. Anal. {\bf 10} (2003) 31--65

\bibitem{BouzDB96}
A.~Bouzouina and S.~De~Bi\`evre, {\it Equipartition
of the eigenfunctions of quantized ergodic maps on the torus}, 
Commun. Math. Phys. {\bf 178} (1996)  83--105

\bibitem{CasMasShep99}
G.~Casati, G.~Maspero and D.~Shepelyansky, {Quantum fractal eigenstates}, 
Physica {\bf D 131} (1999) 311--316

\bibitem{CherMar97} N.~Chernov and R.~Markarian, {\it Ergodic properties of Anosov maps
with rectangular holes}, Boletim Sociedade Brasileira Matematica {\bf 28} (1997) 271--314;
N.~Chernov, R.~Markarian and S.~Troubetzkoy,
{\it Invariant measures for Anosov maps with small holes},
Ergod. Th. Dyn. Sys. {\bf 20} (2000) 1007--1044

\bibitem{CdV85} 
Y.~Colin~de~Verdi\`ere, 
{\it Ergodicit\'e et fonctions propres du laplacien}, 
Commun. Math. Phys. {\bf 102} (1985) 497--502

\bibitem{DENW06}
M.~Degli~Esposti, S.~Nonnenmacher and B.~Winn, 
{\em Quantum Variance and Ergodicity for the baker's map},
Commun. Math. Phys. {\bf 263} (2006) 325--352

\bibitem{DemYou06}
M.~F.~Demers and L.-S.~Young, {\em Escape rates and conditionally invariant
measures}, Nonlinearity {\bf 19} (2006) 377--397

\bibitem{DSZ04}
N.~Dencker, J.~Sj\"ostrand and M.~Zworski, {\em Pseudospectra of semiclassical
(pseudo-)differential operators},  Comm. Pure Appl. Math. {\bf 57} (2004) 384--415

\bibitem{ErmSara06}
L.~Ermann and  M.~Saraceno, {\it Generalized Quantum Baker Maps as 
perturbations of a simple kernel},
Phys. Rev. {\bf E 74} (2006) 046205

\bibitem{FNdB03}
F.~Faure, S.~Nonnenmacher and S.~De~Bi\`evre,
{\it Scarred eigenstates for quantum cat maps of minimal periods},
Commun. Math. Phys. {\bf 239} (2003) 449--492

\bibitem{GaspRice89}
P. Gaspard and S.A. Rice, {\em Scattering from a classically chaotic repellor},
 J. Chem. Phys. {\bf 90} (1989) 2225--2241; 
ibid, {\em  Semiclassical quantization of the
scattering from a classical chaotic repellor},
J. Chem. Phys. {\bf 90} (1989) 2242--2254;
A.~Wirzba, 
{\it Quantum Mechanics and Semiclassics of Hyperbolic n-Disk Scattering Systems},
Physics Reports {\bf 309} (1999) 1--116

\bibitem{GerLei93} 
P.~G\'erard and E.~Leichtnam, 
{\it Ergodic properties of eigenfunctions for the Dirichlet problem}, 
Duke Math. J. {\bf 71} (1993) 559--607

\bibitem{GLZ04} L. Guillop\'e, K. Lin, and M. Zworski, 
{\em The Selberg zeta function for convex co-compact Schottky groups,}
Comm. Math. Phys, {\bf 245} (2004) 149--176

\bibitem{HJPPS86}
T.C.~Halsey, M.H.~Jensen, L.P.~Kadanoff, I.~Procaccia and B.I.~Shraiman, 
{\it Fractal measures and their singularities: The characterization of
strange sets}, Phys. Rev. {\bf A 33} (1986) 1141--1151

\bibitem{HelMarRob87} 
B.~Helffer, A.~Martinez and D.~Robert, {\it Ergodicit\'e et limite semi-classique}, 
Commun. Math. Phys. {\bf 109} (1987) 313--326

\bibitem{Kelmer05}
D.~Kelmer, {\it Arithmetic quantum unique ergodicity for symplectic linear
maps of the multidimensional torus}, to appear in Ann. of Math., \texttt{math-ph/0510079};
{\it ibid}, {\it Scarring on invariant manifolds for perturbed quantized hyperbolic toral automorphisms},
preprint, \texttt{math-ph/0607033}

\bibitem{KNNS07}
J.P.~Keating, S.~Nonnenmacher, M.~Novaes, and M.~Sieber,
{\em On the resonance eigenstates of an open quantum baker map}, 
in preparation.

\bibitem{KNPS06}
J.P.~Keating, M.~Novaes, S.D.~Prado and M.~Sieber,
{\it Semiclassical structure of chaotic resonance eigenfunctions}, 
Phys. Rev. Lett. {\bf 97} (2006) 150406 

\bibitem{KurRud00} 
P.~Kurlberg and Z.~Rudnick, {\it Hecke theory and
equidistribution for the quantization of linear maps of the torus}, 
Duke Math. J. {\bf 103} (2000) 47--77

\bibitem{Leben06}
M.~Lebental, J.-S.~Lauret, J.~Zyss, C.~Schmit and E.~Bogomolny, 
{\it Directional emission of stadium-shaped microlasers}, Phys. Rev. {\bf A 75}
(2007) 033806

\bibitem{LeforWyatt83}
C.~Leforestier and R.E.~Wyatt, {\it Optical potential for laser induced
dissociation}, J. Chem. Phys. {\bf 78} (1983) 2334--2344

\bibitem{Lin02}
K.~Lin, {\it Numerical study of quantum resonances in          
chaotic scattering}, J. Comp. Phys. {\bf 176} (2002) 295--329;
K.~Lin and M.~Zworski, {\it Quantum resonances in          
chaotic scattering}, Chem. Phys. Lett. {\bf 355} (2002) 201--205

\bibitem{Linden06}
E.~Lindenstrauss, {\it Invariant measures and arithmetic quantum unique ergodicity}, 
Annals of Math. {\bf 163} (2006) 165-219

\bibitem{LSZ03}
W.~Lu, S.~Sridhar, and M.~Zworski,
{\em Fractal Weyl laws for chaotic open systems}, 
Phys. Rev. Lett. {\bf 91} (2003) 154101

\bibitem{Mark96}
R.~Markarian, A.~Lopes,
{\it Open billiards: invariant and conditionally invariant probabilities on Cantor sets},
SIAM J. Appl. Math. {\bf 56} (1996) 651--680

\bibitem{MarOK05}
J.~Marklof and S.~O'Keefe, {\em Weyl's law and quantum ergodicity for maps with 
divided phase space}, Nonlinearity {\bf 18} (2005) 277--304

\bibitem{MeenakLaksh04} N.~Meenakshisundaram and A.~Lakshminarayan,
{\it Multifractal eigenstates of quantum chaos and the Thue-Morse sequence},
Phys. Rev. {\bf E 71} (2005) 065303;
ibid, {\it Using the Hadamard and related transforms for simplifying 
the spectrum of the quantum baker's map}, J. Phys. {\bf A 39} (2006) 11205-11216

\bibitem{NonZw05}
S.~Nonnenmacher and M.~Zworski, {\em Fractal Weyl laws in discrete models
of chaotic scattering},
J. Phys. {\bf A 38} (2005) 10683--10702 (special issue on ``Trends in quantum
chaotic scattering''); S.~Nonnenmacher, {\it Fractal Weyl law for open chaotic maps}, in 
{\it Mathematical physics of quantum mechanics}, J.~Asch and A.~Joye Eds.,
Lect. Notes in Physics {\bf 690}, Springer, Berlin, 2006.

\bibitem{NonZw06}
S.~Nonnenmacher and M.~Zworski, 
{\em Distribution of resonances for open quantum maps}, 
Commun. Math. Phys. {\bf 269} (2007) 311--365

\bibitem{NonZw-gap}
S.~Nonnenmacher and M.~Zworski, 
{\em Quantum decay rates in chaotic scattering}, talk given at 
\'Ecole Polytechnique, Palaiseau, France (May 2006), 
\texttt{http://math.berkeley.edu/$\sim$zworski/nzX.ps.gz}

\bibitem{RudSar94}
Z.~Rudnick and P.~Sarnak, {\em The behaviour of eigenstates of arithmetic
hyperbolic manifolds}, Commun. Math. Phys. {\bf 161} (1994) 195--213

\bibitem{Sar90}
M.~Saraceno, {\em Classical structures in the quantized baker transformation}
Ann. Phys. (NY) {\bf 199} (1990) 37--60

\bibitem{SaVa96} 
M.~Saraceno and R.O.~Vallejos, {\it The quantized D-transformation}, 
Chaos {\bf 6} (1996) 193--199

\bibitem{SchaCav00}
R.~Schack and C.M.~Caves, {\it Shifts on a finite qubit string: 
a class of quantum baker's maps},
Appl. Algebra Engrg. Comm. Comput. {\bf 10}, 305--310 (2000)

\bibitem{Schni74} 
A.~Schnirelman, {\it Ergodic properties of eigenfunctions}, 
Usp. Math. Nauk. {\bf 29} (1974) 181--182

\bibitem{SchoTwo04}
H.~Schomerus and J.~Tworzyd{\l}o, {\it Quantum-to-classical crossover of
quasi-bound states in open quantum systems}, 
Phys. Rev. Lett. {\bf 93} (2004) 154102

\bibitem{Roman-PhD}
R.~Schubert, {\it Semiclassical localization in phase space},
PhD thesis, University of Ulm, 2001.

\bibitem{SeidMill92}
T.~Seideman and W.H.~Miller, {\it Calculation of the cumulative reaction
probability via a discrete variable representation with absorbing boundary
conditions}, J.~Chem.~Phys. {\bf 96} (1992) 4412--4422

\bibitem{Sjo90}
J. Sj\"ostrand,  {\em
Geometric bounds on the density of resonances for semiclassical problems},
{ Duke Math. J.}, {\bf 60} (1990) 1--57

\bibitem{SjoZw05} J. Sj\"ostrand and M. Zworski,
{\em Fractal upper
bounds on the density of semiclassical resonances},
to appear in Duke Math. J., \texttt{math.SP/0506307}

\bibitem{Stef05}
P.~Stefanov, {\it Approximating resonances with the Complex Absorbing 
Potential Method},  Commun. PDE {\bf 30} (2005) 1843--1862

\bibitem{TangZw98}
S.-H.~Tang and M.~Zworski, {\it From quasimodes to resonances},
Math. Res. Lett. {\bf 5} (1998) 261--272;
P.~Stefanov, {\it Quasimodes and resonances: sharp lower bounds},
Duke Math. J. {\bf 99} (1999) 75--92

\bibitem{TraSco02}
M.~M.~Tracy and A.~J.~Scott, {\it The classical limit for a class of 
quantum baker's maps}, J. Phys. {\bf A 35} (2002) 8341--8360

\bibitem{Zel87}
S.~Zelditch, 
{\it Uniform distribution of the eigenfunctions on compact hyperbolic surfaces},
Duke Math. J. {\bf 55} (1987) 919--941

\bibitem{Zel96}
S.~Zelditch, {\em Quantum ergodicity of $C^*$ dynamical systems},
Commun. Math. Phys {\bf 177} (1996) 507--528

\bibitem{ZelZwo96} S. Zelditch and M. Zworski, 
{\it Ergodicity of eigenfunctions for Ergodic Billiards}, 
Commun. Math. Phys {\bf 175} (1996) 673--682

\bibitem{Zw99} M. Zworski, {\em Dimension of the limit set and the density of
resonances for convex co-compact Riemann surfaces}, Inv. Math. {\bf 136} (1999)
353--409
\end{thebibliography}
